\documentclass[aps,prb,reprint,superscriptaddress]{revtex4-1}
\usepackage{graphicx}

\usepackage{color}
\usepackage{enumerate}
\usepackage{amsmath}\usepackage{amssymb}\usepackage{stmaryrd}
\usepackage{multirow}
\usepackage{float}
\usepackage{longtable,booktabs}
\usepackage{color}\usepackage{enumerate}
\usepackage{amsmath}\usepackage{amssymb}\usepackage{stmaryrd}
\usepackage{multirow}
\usepackage{float}
\usepackage[toc,page]{appendix}
\usepackage{longtable,booktabs}
 \usepackage{epsfig}
\usepackage{graphicx} 
\usepackage{soul}
\usepackage{bm}
\usepackage{braket}
\usepackage{array}
\usepackage[normalem]{ulem}

\usepackage{feynmp-auto}
\usepackage{psfrag} 
\usepackage{slashed}
\usepackage{amsmath,amssymb} 
\usepackage{colordvi} 
\usepackage{hyperref}
\usepackage{setspace}
\usepackage[all]{hypcap}
\usepackage{subfigure}

\definecolor{darkblue}{rgb}{0.1,0.2,0.6}
\definecolor{darkred}{rgb}{0.8,0.1,0.2}
\hypersetup{
    bookmarks=true,         
    unicode=false,          
    pdftoolbar=true,        
    pdfmenubar=true,        
    pdffitwindow=false,     
    pdfstartview={FitH},    
    pdftitle={My title},    
    pdfauthor={Author},     
    pdfsubject={Subject},   
    pdfcreator={Creator},   
    pdfproducer={Producer}, 
    pdfkeywords={keyword1} {key2} {key3}, 
    pdfnewwindow=true,      
    colorlinks=true,       
    linkcolor=blue,          
    citecolor=darkred,        
    filecolor=magenta,      
    urlcolor=darkblue           
}
\def\be{\begin{equation}}
\def\bea{\begin{eqnarray}}
\def\eea{\end{eqnarray}}
\def\ee{\end{equation}}
\def\del{\partial}

\def\Tr{{\rm Tr }}

\def\kv{\textbf{k}}
\def\Z{\mathbb{Z}}

\def\be{\begin{equation}}
\def\bea{\begin{eqnarray}}
\def\eea{\end{eqnarray}}
\def\ee{\end{equation}}
\def\del{\partial}

\def\r{{\bf r}}
\renewcommand{\[}{\left [}
\renewcommand{\]}{\right ]}
\renewcommand{\(}{\left (}
\renewcommand{\)}{\right )}

\begin{document}

\title{Topological crystalline superconductivity and second-order topological superconductivity in {nodal-loop materials}}
\author{Hassan Shapourian}
\affiliation{James Franck Institute and Kadanoff Center for Theoretical Physics, University of Chicago, Illinois 60637, USA}
\author{Yuxuan Wang}
\email{yxwang@illinois.edu}
\affiliation{Department of Physics and Institute for Condensed Matter Theory, University of Illinois at Urbana-Champaign, Urbana, Illinois 61801, USA}
\author{Shinsei Ryu}
\affiliation{James Franck Institute and Kadanoff Center for Theoretical Physics, University of Chicago, Illinois 60637, USA}
\date{\today}

\begin{abstract}
We study the intrinsic fully-gapped odd-parity superconducting order in doped nodal-loop materials with a torus-shaped Fermi surface. We show that the mirror symmetry, which protects the nodal loop in the normal state, also protects the superconducting state as a topological crystalline superconductor. As a result, the surfaces preserving the mirror symmetry host gapless Majorana cones. Moreover, for a Weyl loop system (two-fold degenerate at the nodal loop), the surfaces that break the mirror symmetry (those parallel to the bulk nodal loop) contribute a Chern (winding) number to the quasi-two-dimensional system in a slab geometry, which leads to a quantized thermal Hall effect and a single Majorana zero mode bound at a vortex line penetrating the system.
This Chern number can be viewed as a higher-order topological invariant, which supports hinge modes in a cubic sample when mirror symmetry is broken.
For a Dirac loop system (four-fold degenerate at the nodal loop), the fully gapped odd-parity state can be either time-reversal symmetry-breaking or symmetric, similar to the $A$- and $B$- phases of $^3$He. In a slab geometry, the $A$-phase has a Chern number two, while the $B$-phase carries a nontrivial $\mathbb{Z}_2$ invariant. We discuss the experimental relevance of our results to nodal-loop materials such as CaAgAs.
\end{abstract}

\maketitle

\section{Introduction}

Over the past few years, the study of topological band structures has been extended from gapped systems to gapless systems, which exhibit topologically protected gap closing nodes. Other than the well-known Weyl and Dirac semimetals (For reviews, see e.g., Refs.\ \onlinecite{armitage-review, yan-review}), where gapless points form isolated points in momentum space, another particularly interesting case is when the gapless points form a closed loop~\cite{Carter2012,Chen2015,Schaffer2015,Kim2015,Mullen2015,Zeng2015,Weng2015,Yu2015,Xie2015,Bian2016,nini-2017,nini2-2017,dessau-2017}. These so-called nodal-loop semimetals have recently been proposed to exist in e.g., Ca$_3$P$_2$~\cite{Xie2015}, TlTaSe$_2$~\cite{Bian2016}, CaAgP and CaAgAs~\cite{Yamakage2016}. In the case of CaAgAs, magneto-transport properties~\cite{Okamoto2016,nini-2017} and direct measurement of angle-resolved photoemission spectroscopy (APPES)~\cite{Takane,Nayak} show evidence of a single nodal loop near the Fermi level. Depending on the band structure, the nodal loop can either be two-fold degenerate or four-fold degenerate, commonly referred to as Weyl loop and Dirac loop, respectively. 
The nodal loop can be thought of
as
sources of Berry curvature in momentum space, thus a closed loop surrounding the nodal line has Berry phase of $\pi$. As a result of the nontrivial topology, there exist degenerate two-dimensional (2D) energy bands that are localized on surfaces parallel to the nodal loop whose momentum range is confined by the projection of the nodal loop, also known as the ``drumhead bands"~\cite{Volovik2011,Burkov2011,Chan2016,Ramamurthy2017}. Another remarkable feature of the nodal-ring band structure is the torus-shaped Fermi surface (FS) if the Fermi level is not exactly at the nodal point (neutrality). There is currently great theoretical and experimental interest in searching and exploring exotic properties in these materials~\cite{Rhim2015,Liu2017}.

While most studies on nodal-loop materials focus on the band structure in the free-electron limit, a natural extension is to consider interaction effects~\cite{Nandkishore2016,Wang-Ye2016,Liu2017,Roy2017,wang-nandkishore2-2017}. One particularly interesting direction is superconducting orders which develop at low temperatures~\cite{Guan-2016}. For the Weyl loop case, it was found that due to the nontrivial (pseudo)-spin texture on the torus FS, the leading superconducting instability is toward a chiral $p$-wave pairing~\cite{Nandkishore2016,Sur-Nandkishore2016}, which leads to a full gap.  Given the nontrivial band structure and the $p_x + i p_y$ order, it is interesting to ask what is the topology of the nodal loop superconductor. Via arguments in analogy with the 2D $p_x + i p_y$ superconductivity and numerical calculations, a previous work by one of us~\cite{Wang-Nandkishore2017} suggested that the $p$-wave order induces topological superconductivity on the surface drumhead bands. However, unlike a true 2D $p_x + i p_y$ superconductor, which has a trivial phase and topological phase depending on chemical potential,  it was argued that the drumhead superconductor is always in the topological phase.
Thus nodal-loop materials provide a new direction for the search of novel topological superconductors. Nonetheless, a full understanding of the topological characterization of the chiral $p$-wave order is still lacking. 
In particular, we note that the chiral $p$-wave order, which breaks time-reversal
symmetry, belongs to class D of the Altland-Zirnbauer symmetry
classes~\cite{AZ_1997}. However, in three-dimensions, the topological classification of class D
is empty (always trivial). This then brings into question how the surface
topological supercondoctor
fits into the ``ten-fold way" classification table~\cite{Schnyder_class,Kitaev_class,Ryu_class}.
 In this paper we provide such a systematic analysis.  

The main results of our work are summarized as follows. For a Weyl loop semimetal,
crystalline symmetries, for example a mirror symmetry $\mathcal{R}_z$ in the perpendicular direction to
the nodal loop (hereafter referred to as the $z$-direction) can be invoked to make the nodal loop 
stable~\cite{Burkov2011}. Importantly, we find that the \emph{same} mirror reflection symmetry also ensures the nontrivial topology in the superconducting state. Using the classification table~\cite{Chiu2013,Morimoto2013,Shiozaki2014,Chiu_RMP},
we find that the Weyl loop superconductor belongs to class ``D$+{\cal R}_+$", and it has a $\mathbb{Z}$ classification characterized by a mirror Chern number $\nu_{k_z=0,\pi}$. 
We obtain $\nu_{k_z=0}=1$ for the $p$-wave state, which implies that the Weyl loop superconductor is in a topological phase with the $\Z$ topological index equals one. 
The non-trivial topology of such a phase leads to gapless Majorana cones on mirror symmetric open surfaces (say $yz$ and $xz$ planes).

On the other hand, the two $xy$ surfaces parallel to the nodal loop break mirror symmetry and are gapped. We show that for a system in a slab geometry along $z$ direction,  each surface contributes a  Chern number of $\pm 1/2$ 
and we verify this numerically~\cite{Essin-Moore}.
We thus predict a quantized thermal Hall effect corresponding to Chern number $C=1$. Moreover, a vortex line penetrating the system in the $z$-direction carries a \emph{single} Majorana mode. 
We study the wave function of the vortex core bound state.   With mirror symmetry, the vortex line Majorana wave function is smeared along the vortex line.
In the presence of a mirror symmetry breaking term $\sim E$, the resulting vortex core Majorana mode becomes localized at either top or bottom surface. We analytically and numerically solve the Bogoliubov-de-Gennes (BdG) equation to verify this. The dependence of the wave function profile on small symmetry breaking effects offers an interesting potential application of controlling Majorana zero modes via applying an electric field (i.e. gating). 

Similarly, one can show that in a cube geometry, with broken mirror symmetry, the $xz$ and $yz$ surfaces are gapped, and carry non-zero $\pm1/2$ Chern numbers. Hence, there are chiral modes formed as domain-wall states between surfaces with opposite Chern numbers. In this case, the modes are localized  at the top or bottom \emph{hinges}~\cite{Hughes1, Hughes2, Fang, Neupert, Luka, neupert_new} of the system. The Chern number $C=1$ can be viewed as a higher-order topological invariant {identified in Ref.\ \onlinecite{Luka}}, and in this sense the nodal loop superconductor is a higher-order topological superconductor with mirror symmetry.
We show that this Chern number $C=1$ is stable against a mirror symmetry breaking field $E$, as long as the Fermi surface remains torus-shaped, and obtain a global phase diagram as a function of chemical potential $\mu$ and electric field $E$ (shown in Fig.~\ref{fig:phasediag2}), and specify the topological regions. 


We extend our analysis to Dirac-loop superconductors, which is at this stage more relevant to experimental materials (for example CaAgP and CaAgAs). In the absence of spin-orbit coupling, Dirac-loop semimetal can be thought of as two copies of Weyl-loop semimetals with spin-up and spin-down respectively. With a proper pairing mechanism, for example via a short-range repulsive interaction~\cite{Sur-Nandkishore2016}, one may still expect the leading instability to be towards  spin-triplet and orbit-triplet $p$-wave states, which is fully gapped. Similar to the superfluid $^3$He, we discuss the topology of the time-reversal breaking phase (denoted as $A$-phase) and time-reversal invariant phase ($B$-phase).
The $A$-phase of the Dirac loop superconductor fits in the same topological classification of class D$+{\cal R}_{+}$, now with $\nu_{k_z=0}=2$ resulting in two Majorana cones on each mirror symmetric surface. For a slab geometry parallel to the nodal loop, there is a total  Chern number $C=2$, which leads to a quantized thermal Hall effect. 
 The $B$-phase can be thought of as a $p_x + i p_y$ order for one spin species and $p_x - i p_y$ for the other, which are related by time-reversal.  Topologically it fits in class DIII$+{\cal R}_{++}$, which has a $\mathbb{Z}_2$ classification, and we show that our $B$-phase indeed is topological. In a quasi-2D slab geometry, it carries a $\mathbb{Z}_2$ topological invariant (which can also be viewed as a second-order topological invariant~\cite{Luka}). Hence, the vortex-line core binds two Majorana zero modes which are related and protected by time-reversal symmetry. 
  
The rest of the paper is organized as follows. In Sec.\ \ref{sec:weyl} we focus on the topological crystalline superconductivity in doped Weyl loop materials. We first address the structure of the superconducting order, and then analyze its topology both in the bulk and for a quasi-2D slab geometry. In Sec.\ \ref{sec:dirac} we analyze the topological crystalline superconductivity in doped Dirac loop materials. We discuss both the $A$-phase and the $B$-phase and their topology, as well as the effect of spin-orbit coupling. In Sec.\ \ref{sec:conclusions} we present the conclusions and discuss the relevance of our results to experiments.

\section{chiral $p$-wave Weyl loop superconductors}\label{sec:weyl}

\subsection{Superconductivity in a doped Weyl loop semimetal}
We first consider the case where the nodal loop is formed by two bands crossing along a loop in the Brillouin zone. 
To be more specific, let us consider the following two-band Hamiltonian as a minimal model 
\begin{align} \label{eq:Hcont}
{H}=\sum_\kv c_\kv^\dag \left[\frac{k_x^2+k_y^2-k_F^2}{2m}\sigma_1+ v_z k_z\sigma_2 -\mu \right] c_\kv.
\end{align}
where $c_\kv^\dag=(c_{\kv1}^\dag,c_{\kv2}^\dag)$ is composed of two fermion operators associated with  opposite (pseudo)-spins (which may be real spin or some orbital degrees of freedom) and $\mu$ is the chemical potential. This model has a loop of Weyl nodes along the circle $k_x^2+k_y^2=k_F^2$ located in the $k_z=0$ plane and the Fermi surface is torus-shaped at non-zero chemical potential $\mu$. 
The nodal loop is protected by a mirror reflection symmetry ${{\cal R}_{z}}$ where the symmetry transformation is given by
\begin{align} \label{eq:reflection}
{{\cal R}_{z}} c^\dag_{\kv} {{\cal R}_{z}}^{-1}= R_z c^\dag_{{{\cal R}_{z}}\kv} 
\end{align}
in which ${{\cal R}_{z}}\kv=(k_x,k_y,-k_z)$ and $R_z=\sigma_1$.

\begin{figure}
\centering
\includegraphics[scale=.8]{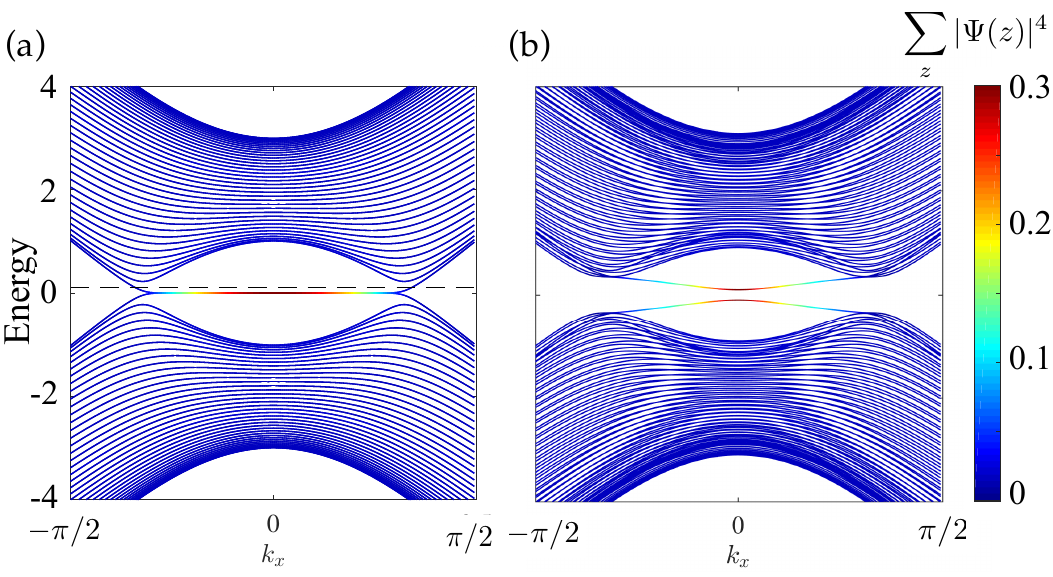}
\caption{\label{fig:bands} Low energy band structure of a slab of the nodal-loop semimetal in the normal phase (\ref{eq:nlsm}) and (b) superconducting phase (\ref{eq:bdg}). The color code shows the inverse participation ratio (\ref{eq:IPR}) to distinguish topological surface modes localized near top and bottom surfaces from the delocalized bulk states. The dashed line in (a) shows the chemical potential  $\mu=0.2$.  The slab consists of $40$ layers and $t_3=0$. For (b), we set $\Delta=0.4$.}
\end{figure}

For a finite system, on the surfaces parallel to the nodal loop ($xy$ surfaces), there exist the so-called ``drumhead" surface states. 
These surface states can be derived as the  domain wall zero-mode solutions of an interface with a trivial insulator ${\cal H}= M\sigma_1$ ($M>0$). The resulting bands are dispersionless 
\be 
{\cal H}(\kv_\parallel)=-\mu
\ee
 and spread within the circle $k_x^2+k_y^2=k_F^2$ (that is the $xy$-plane projection of the bulk nodal loop), where they join the bulk states [see Fig.~\ref{fig:bands}(a)]. The surface states are found to be pseudospin polarized, i.e., $\sigma_3\ket{\Psi}=\pm \ket{\Psi}$ on the top and bottom surfaces, respectively. The top and bottom surface states are degenerate, which is also protected by the mirror symmetry ${\cal R}_{z}$.

Due to the underlying nodal ring, the torus Fermi surface has a nontrivial spin texture. As a result, it is a natural host of unconventional superconducting orders~\cite{Nandkishore2016,Sur-Nandkishore2016,Wang-Nandkishore2017}. It was found in Ref.\ \onlinecite{Wang-Nandkishore2017} that even the usual $s$-wave order exhibits line nodes on the Fermi surface. For a repulsive interaction, the leading superconducting instability is towards $p$-wave order~\cite{Sur-Nandkishore2016} of the form $H_{\Delta}\sim c_{\kv}({\bf d_\kv}\cdot\vec\sigma) i\sigma_2 c_{-\kv}+{\rm H.c}$, where ${\bf d_\kv}$ is odd in $\kv$. To fully gap the Fermi surface~\cite{Nandkishore2016,Sur-Nandkishore2016}, the $\bf d$-vector points in the $y$ direction, and the $p$-wave order takes the chiral form of
\begin{align}\label{eq:p+ip}
{H}_\Delta=\sum_{\kv,\sigma}\left[ \Delta (k_x+i k_y) c_{\kv\sigma}^\dag c_{-\kv\sigma}^\dag+ \text{H.c.}\right].
\end{align}
Consequently, one finds that the effective surface Hamiltonian is given by (see Appendix \ref{app:surface})~\cite{Wang-Nandkishore2017}
\begin{align} \label{eq:surface}
{\cal H}(\kv_\parallel)= \Delta (k_x \tau_x + k_y \tau_y) - \mu \tau_z,
\end{align}
where $\tau$ matrices are defined in the Nambu space. This resembles the low-energy theory of a $p_x+i p_y$ superconductor~\cite{ReadGreen2000}, and {is confirmed by the low-energy dispersion shown in Fig.~\ref{fig:bands}(b)}.  

Next, we characterize the topological properties of such superconducting phase and in particular the surface superconductivity. To better diagnose the topology of the bulk and surface states, we use a simple microscopic lattice model described by the momentum space  Hamiltonian $H_\text{sm}=\sum_\kv c_\kv^\dag {\cal H}_\text{sm}(\kv) c_\kv$, 
\begin{align} \label{eq:nlsm}
{\cal H}_\text{sm} =&  (6 - t_1 -2 \cos k_x - 2 \cos k_y -2 \cos k_z) \sigma_1\nonumber \\
& + 2 t_2  \sin (k_z) \sigma_2 - \mu \nonumber \\
& +t_3(2-\cos k_x-\cos k_y),
\end{align}
which reproduces the low energy theory \eqref{eq:Hcont} near $\kv=0$. Here, $t_2$ and $t_1$ are a hopping amplitude and a mass parameter, respectively (for simplicity, we take $t_1=t_2=1$ unless otherwise stated).    {We have also included the $t_3$ term in the Hamiltonian to add a  quadratic dispersion to the flat surface drumhead bands, since it is allowed by the reflection symmetry $\mathcal{R}_z$.}

\begin{figure}
\centering
\includegraphics[scale=1.3]{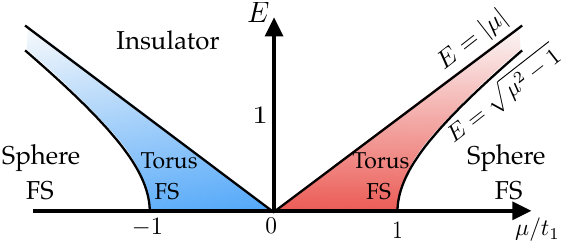}
\caption{\label{fig:phasediag} Topology of the Fermi surface (FS) of the nodal-loop semimetal model~(\ref{eq:nlsm}) as a function of chemical potential $\mu$ in the presence of reflection symmetry breaking term $E$. The insulating region refers to absence of any Fermi surface. This phase diagram is plotted for $t_3=0$. For $t_3\neq 0$, there will be minor modifications to the phase boundaries, while the relative locations of phases do not change.}
\end{figure}

 At a finite chemical potential $|\mu|<t_1$, the bulk Fermi surface is torus-shaped. 
 A typical band structure as a function of $k_x$ at $k_y=0$ for a slab (finite in $z$ direction) of the nodal-loop semimetal (\ref{eq:nlsm}) is plotted in Fig.~\ref{fig:bands}(a). The drumhead bands appear in the middle of the spectrum and their corresponding wave functions are localized near the top and bottom surfaces.  To illustrate  the localization of eigenstates, we compute the inverse participation ratio (IPR)~\cite{IPR_Thouless},
\begin{align} \label{eq:IPR}
\text{IPR}_\Psi=\sum_z |\Psi(z)|^4
\end{align}
where $|\Psi(z)|^2=\sum_{x,y} |\Psi(\textbf{r})|^2$ is the squared amplitude of the wave function in the $z$-th layer and it is assumed that the wave function is normalized $\sum_\textbf{r}|\Psi(\textbf{r})|^2=1$. The IPR is known from the context of Anderson localization and indicates how localized a state is in the following way: If a wave function is localized on a single site, IPR will be close to one and independent of system size. If a wave function is uniformly delocalized across the system, then IPR will go to zero as $1/L_z$ where $L_z$ is the number of layers. In the case of problem at hand, $\text{IPR}=1/2$ in the limit of infinite layers indicates localization near the top and bottom surfaces.
 The IPR of eigenstates are shown as a color code in Fig.~\ref{fig:bands}(a), which clearly indicates that the bulk states are delocalized throughout the sample while the surface bands are very localized.

 In order to investigate the stability of topological properties in the absence of the reflection symmetry (\ref{eq:reflection}), it is instructive to also introduce a symmetry breaking term
\begin{align} \label{eq:Efield}
{\cal H}_E= E  \sigma_3,
\end{align}
 which effectively acts like a mirror-breaking electric field along the $z$ direction. It gaps out the nodal line, and induces an energy imbalance between the top and bottom surface drumhead bands (which are labeled by $\sigma_3=\pm 1$).
{The topology of the bulk Fermi surface in our model (\ref{eq:nlsm}) depends on the chemical potential $\mu$ as well as the {electric field} $E$. This dependence is represented by a two-variable ``phase diagram", shown in Fig.~\ref{fig:phasediag}. For a negligible $t_3$ (the inclusion of which does not change the global structure of the phase diagram, as we have verified),
when the chemical potential is small $|\mu|<t_1$ the bulk Fermi surface is a torus, but for larger values $|\mu|>t_1$ the Fermi surface changes topology and becomes topologically {equivalent to} a sphere (no longer has a ``central hole"). Furthermore, as we add the electric field term (\ref{eq:Efield}), the nodal loop opens a gap and we get an insulating phase whenever the chemical potential is in the gap $|\mu|<E$. However, as we increase $\mu$ we recover a toroidal Fermi surface in the regime $E<|\mu|<\sqrt{1+E^2}$ and eventually, we obtain a spherical Fermi surface for larger values $|\mu|>\sqrt{1+E^2}$. The two regions with (electron-like and hole like) torus FS's are the regime of interest throughout this work.}


As mentioned, in this paper we are interested in the regime where the Fermi surface is a torus and the dominant superconductivity is a chiral $p$-wave pairing. The corresponding term in the lattice Hamiltonian for $p_x+ i p_y$ superconductivity reads as
$H_\Delta=\Delta (\sin k_x + i\sin k_y) c_{\kv\sigma}^\dag c_{-\kv\sigma}^\dag+ \text{H.c.}$
As a result, the full BdG Hamiltonian ${ H}={ H}_\text{sm}+ { H}_\Delta=\sum_\kv \psi_\kv^\dag {\cal H}(\kv) \psi_\kv$ can be written as 
\begin{align} \label{eq:bdg}
{\cal H} ({\bf k})= & (6 - t_1 -2 \cos k_x - 2 \cos k_y -2 \cos k_z) \tau_z \sigma_1  \nonumber \\
&+ 2 t_2   \sin (k_z) \tau_z \sigma_2- \mu  \tau_z \sigma_0 
 \nonumber \\
 &+ t_3  (2-\cos k_x -\cos k_y)\tau_z \sigma_0 \nonumber \\
 &+ \Delta  (\tau_x \sin k_x + \tau_y \sin k_y) \sigma_0,
  \end{align}
  in the basis
$\psi_\kv^\dag=(c_{\kv1}^\dag,c_{\kv2}^\dag,c_{-\kv1},c_{-\kv2})$
where $\tau_i$ are Pauli matrices acting on particle-hole degrees of freedom. Here we have used the shorthand $\tau_\alpha \sigma_j$ for the Kronecker product $\tau_\alpha \otimes \sigma_j$.
The above Hamiltonian is symmetric under the mirror  reflection {$\mathcal{R}_z$} in (\ref{eq:reflection}). 
The BdG Hamiltonian has a particle-hole symmetry by construction,  $\mathcal{H} ({\bf k})= - C \mathcal{H} (-{\bf k}) C^{-1}$, where $C=\tau_x {\cal K}$ and ${\cal K}$ is the complex conjugation operator.

Typical eigenvalues of the BdG Hamiltonian (\ref{eq:bdg}) as a function of $k_x$ at $k_y=0$ for a slab (finite in $z$ direction) is shown in Fig.~\ref{fig:bands}(b). As it is evident from this plot, the surface states can be clearly distinguished from the bulk states by IPR.

\subsection{Topology of the chiral $p$-wave Weyl loop superconductor}

In this section, we discuss the topological properties of the four-band BdG Hamiltonian~(\ref{eq:bdg}), both in the bulk and in a slab geometry parallel to plane of the nodal loop.

\subsubsection{Topology in the bulk}

As we noted earlier, the Hamiltonian \eqref{eq:bdg} with $p_x+ip_y$ superconducting order breaks time-reversal symmetry and belongs to class D of the Altland-Zirnbauer symmetry classes, which has a particle-hole symmetry. However, if we only consider the particle-hole symmetry, according to the classification table~\cite{Schnyder_class,Kitaev_class,Ryu_class} in three dimensions there is only a trivial phase. {To characterize the topology}, the mirror reflection symmetry (\ref{eq:reflection}) is crucial. Equipped with ${{\cal R}_{z}}$, our system falls into in class D$+{{\cal R}}_+$ topological crystalline superconductors (following the notation of Ref.~\cite{Chiu2013}, ${{\cal R}}_\pm$ indicates whether the reflection operator $R_z$ commutes or anticommutes with the particle-hole operator $C=\tau_x {\cal K}$). 
The topological classification of the symmetry class D$+{{\cal R}}_+$ is $M\Z$ characterized by a strong topological index from the mirror Chern numbers~\cite{fu-teo-kane, Chiu2013}. The mirror Chern number $\nu_{k_z}$ is defined~\cite{fu-teo-kane} as the Chern number at mirror invariant planes $k_z=0$ and $k_z=\pi$ for bands with a \emph{given} eigenvalue of the mirror operator ${\cal R}_{z}$. The strong topological index is then obtained by~\cite{Chiu2013},
 \begin{align}
 N_{M\Z}= \text{sgn} (\nu_0-\nu_\pi) (|\nu_0|-|\nu_\pi|).
 \end{align}
For our model~(\ref{eq:bdg}), the corresponding Hamiltonians 
 at $k_z=0$ and $\pi$ read as (eigenstate with ${{\cal R}_{z}}=+1$ corresponds to setting $\sigma_1=1$)
 \begin{align}
 {\cal H}_{k_z=0}^{{{\cal R}_{z}}=+1} = & (3-\mu -2 \cos k_x - 2 \cos k_y) \tau_z  \nonumber \\
 &+ \Delta  (\tau_x \sin k_x + \tau_y \sin k_y),  \\ 
 {\cal H}_{k_z=\pi}^{{{\cal R}_{z}}=+1} = & (7-\mu -2 \cos k_x - 2 \cos k_y) \tau_z  \nonumber \\
 &+ \Delta  (\tau_x \sin k_x + \tau_y \sin k_y) ,   
  \end{align}
  which, for the regime of interest $|\mu|<t_1$, is identical to topological and trivial phases of a 2D $p_x + i p_y$ superconductor, respectively. Hence, $|\nu_0|=1$ and $\nu_\pi=0$ and the strong topological index is 
  \be
  |N_{M\Z}|=1.
  \ee 
  Thus, the chiral $p$-wave Weyl loop superconductor is a topological crystalline superconductor in class D$+{{\cal R}}_+$.

 This fact in turn implies that every mirror symmetric surface, in the present case $xz$ and $yz$ surfaces, hosts a gapless Majorana cone which is protected by the mirror symmetry. In Appendix \ref{app:surface} we compute the effective low-energy Hamiltonian for states localized at $xz$ and $yz$ surfaces. For example, the effective low-energy Hamiltonian for $xz$ surface states is given by
 \begin{align} \label{eq:sidesurface}
 h({\bf k}) = \Delta k_x \eta_z - 2 t_2 k_z  \eta_y,
 \end{align}
 where $\eta_{y,z}$ are Pauli matrices in the basis spanned by two spinors $\Psi$ with $\sigma_1\tau_x \Psi=\Psi$. A similar conclusion holds for $yz$ surfaces. The gapless low-energy Hamiltonian in the superconducting state may be measured by ARPES experiments.
 
 \subsubsection{Topology in a slab geometry}

 Beyond the non-trivial bulk topology, we are interested in {the topology carried by} the the  drumhead bands {on the top and bottom surfaces in the superconducting state}. 
 These surfaces, which necessarily break the mirror symmetry in $z$-direction, do not host gapless states protected by the symmetry (see Appendix~\ref{app:surface}), 
and the bulk topological index is not sufficient to describe the topology on these boundary surfaces. 

We consider  the model Hamiltonian (\ref{eq:bdg}) in a slab geometry where the system is terminated along the $z$-direction leading to two open surfaces  perpendicular to the $z$-axis. Using the low energy long-wavelength expansion of the lattice model (\ref{eq:bdg}) near $\kv=0$, we arrive at the effective surface Hamiltonian (\ref{eq:surface}) to the leading order in $\mu$.{ Due to an apparent similarity between the Hamiltonian (\ref{eq:surface}) and that of a two-dimensional $p_x+i p_y$ superconductor~\cite{ReadGreen2000} with Chern (winding) number $C=1$, 
 one may expect a nontrivial topology coming from the surfaces.
On the other hand, an important difference is that the surface effective theory does not cover the entire 2D Brillouin zone and is connected to the bulk bands, and therefore cannot be treated separately from the bulk.} 

To provide direct evidence for the topology in analogy to the $p_x+ip_y$ superconductor, we {treat the slab  as a two-dimensional system and} compute the  Chern number using the TKNN formula~\cite{TKNN1982}
\begin{align} \label{eq:chern}
C = \frac{ 2\pi}{i L_x L_y} \sum_{\kv_\|}  \text{Tr}{\big[} {\cal P}_{\kv_\|} \epsilon_{ij} (\partial_i {\cal P}_{\kv_\|}) (\partial_j {\cal P}_{\kv_\|}) {\big]},
\end{align}
in which $\kv_\|=(k_x,k_y)$ refers to the in-plane momentum. $k_z$ is no longer well-defined, and the degree of freedom of motion in $z$ direction is treated as a band index.
${\cal P}_{\kv}=\sum_\nu |u_{\nu\kv}\rangle \langle u_{\nu\kv}|$ is the projection operator onto the occupied states of the Hamiltonian, and $\partial_i = \frac{\partial}{\partial k_i}$.
Remarkably, we find that the  Chern number is $C=-1$ (for $-t_1<\mu<0$) or $C=1$ (for $0<\mu<t_1$). This means that the system in a quasi-2D slab geometry indeed behaves as a $p_x+ip_y$ superconductor. However,   {depending on the sign of the  chemical potential $\mu$}, for a 2D $p_x+ip_y$ superconductor there is a trivial ``strong pairing" phase with $C=0$ and a topological ``weak pairing" phase with $C=1$~\cite{ReadGreen2000,Alicea2012}; here both cases are topological since $C=\pm 1$.

As a result of the  Chern number $C=1$, we predict that the chiral $p$-wave Weyl loop superconductor exhibits a quantized thermal Hall effect  $j_x^T = \sigma^T_{xy} (\partial_y T)$ with~\cite{Volovik,ReadGreen2000}
\be\label{eq:THall}
\sigma_{xy}^T = \pm \frac{(\pi k_B)^2 T}{6h},
\ee
where $j^T$ is the thermal current and $\partial_y T$ is a temperature gradient.

\begin{figure}
\centering
\includegraphics[scale=.87]{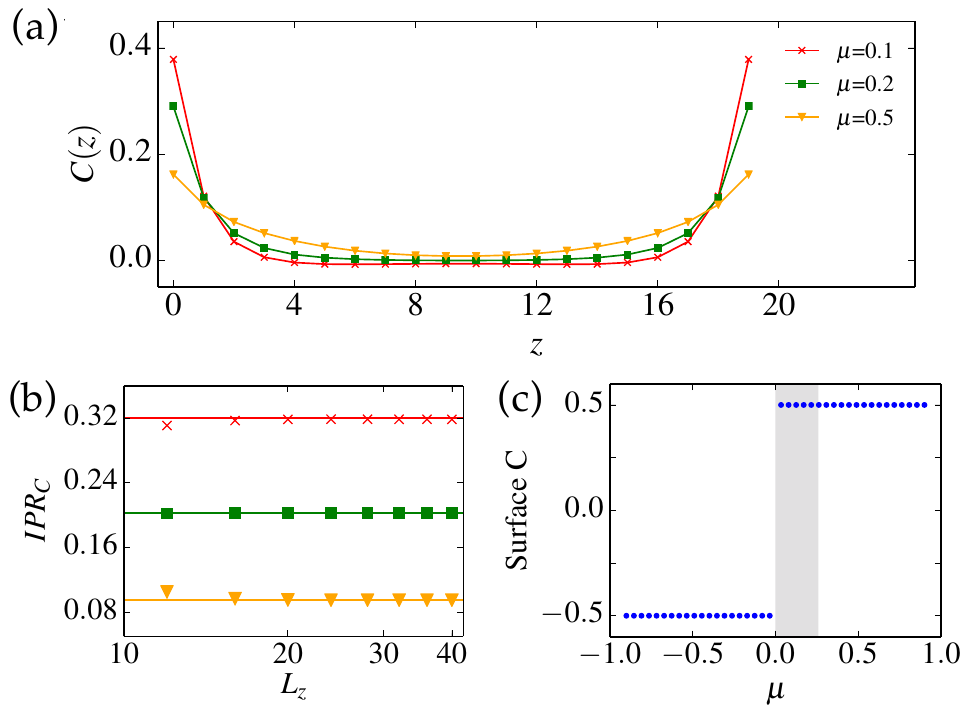}
\caption{\label{fig:Chern} (a) The layer resolved  Chern number (\ref{eq:LRchern}) for various values of chemical potential corresponding to different shapes of Fermi surface (see upper panels of Fig.~\ref{fig:Majorana}). 
(b) The inverse participation ratio for the layer resolved  Chern number $C(z)$, characterizing the localization of the distribution in Panel (a).
(c) The surface  Chern number as computed by $\sum_{z=0}^{L_z/2} C(z)$. The gray area indicates where the chemical potential $\mu$ crosses the surface bands and there is a surface Fermi contour.
The system thickness is $L_z=20$ and we take a $200\times 200$ grid to evaluate the sum over in-plane momenta in Eq.~(\ref{eq:LRchern}). Here, $t_3=0.5$ and $\Delta=0.4$.}
\end{figure}

A natural question here is where the  Chern number ``comes from". 
To address this, we use a more stringent probe to further determine the contribution of bulk states as well as surface states to the  Chern number. The ``layer-resolved"  Chern number~\cite{Essin-Moore,Shapourian2015} is given by
\begin{align}  \label{eq:LRchern}
C(z) = \frac{ 2\pi}{i L^2} \sum_{\kv}  \text{Tr}{\big[} {\cal P}_{\kv} \epsilon_{ij} (\partial_i {\cal P}_{\kv}) |z\rangle \langle z| (\partial_j {\cal P}_{\kv}) {\big]},
\end{align}
where $|z\rangle \langle z|$ is the projection operator onto the $z$-th layer. The results of the layer-resolved  Chern number are plotted in Fig.~\ref{fig:Chern}. It is evident from Fig.~\ref{fig:Chern}(a) that the major contribution comes from the layers near the boundary surfaces. We quantify this observation by computing an inverse participation ratio for $C_z$, defined similarly to \eqref{eq:IPR} as
\begin{align}
\text{IPR}_C=\sum_z |C(z)|^2,
\end{align}
noting that $\sum_z |C(z)|=1$. 
The results are shown in Fig.~\ref{fig:Chern}(b). We see that in the limit $L_z\to \infty$, the IPR$_C$ tends to a constant, {which indicates that distribution of $C_z$ is ``localized"}.  Fig.~\ref{fig:Chern}(c) shows the ``surface" Chern number, obtained by summing over the top (bottom) half number of layers. We conclude that the  Chern number comes equally from the top and bottom surfaces. Physically, for an infinitesimally small $\mu$, the half surface  Chern number is a result of the gapped Majorana (Dirac) cone effective low-energy Hamiltonian \eqref{eq:surface} of the surface states, and similar results are known for topological insulators in the presence of infinitesimal time-reversal symmetry breaking effects~\cite{Qi_3DTI,Essin-Moore}. However, the difference here is that even for a \emph{finite} $\mu$ the surface  Chern number is stable.

The origin of this stable half surface  Chern number is tied with mirror symmetry.
It turns out that the 
topological {gravitational} response theory {of the superconducting state} is  given by a $\theta$-term,
\begin{align}
\mathcal{S}=\frac{\theta}{1536\pi^2}\int \sqrt{g} \epsilon^{cdef}\mathcal{R}^a_{bcd} \mathcal{R}^b_{aef},
\label{Riemann}
\end{align}
{where $\mathcal{R}^a_{bcd}$ is the Riemann  curvature tensor.}
 The form of Eq.~\eqref{Riemann} is the same as that of a 3D time-reversal invariant topological superconductor~\cite{Ryu2012,Nomura2012}, and is a gravitational analog of magnetoelectric effect in topological insulators~\cite{Qi_3DTI}. The $\theta$ angle for a non-interacting band structure can be found by the Berry phase expression~\cite{Qi_3DTI},
\begin{align}  \label{eq:theta}
\theta=\frac{1}{2\pi} \int_{\text{BZ}} d^3 k\ \epsilon_{ijk} \Tr\left[ a_i \partial_j a_k - i \frac{2}{3} a_i a_j a_k \right],
\end{align}
where $a_j^{\mu\nu}=i \bra{u_{\mu\kv}}\partial_j\ket{u_{\nu\kv}}$ is the Berry connection defined in terms of the Bloch functions of occupied bands $\ket{u_{\nu\kv}}$ and $\partial_j=\partial/\partial k_j$. {From the expression \eqref{eq:theta}, we see that in the presence of either mirror reflection symmetry, or time-reversal symmetry, $\theta$ is quantized to be $0$  or $\pi$, since either symmetries send $\theta$ to $-\theta$ and $\theta$ is $2\pi$ periodic. While time-reversal symmetry is usually invoked in the context of topological insulators and superconductors for the quantization of $\theta$, in our model~(\ref{eq:bdg}), such quantization is protected by the mirror reflection symmetry (\ref{eq:reflection}). Via a direct computation we find that 
\be
\theta=\pi.
\ee
By treating the top/bottom surface as a domain wall between $\theta=0$ and $\pi$, it is straightforward to show~\cite{Qi_3DTI,Ryu2012,Nomura2012} that the boundary is described by a gravitational Chern-Simons term with a  Chern number 
\be\label{modulo}
C_{d,u}=\pm\frac{\theta}{2\pi} \rm{~modulo~} 1,
\ee
where $d,u$ denotes bottom and top surfaces respectively.
The ambiguity of the surface  Chern number comes from the fact that $\theta$ is only well-defined modulo $2\pi$.
Therefore, the $\theta$ angle accounts for the anomalous surface  Chern number
$C_{u,d}=\pm 1/2$ in our case -- \emph{even if} the surface Majorana cone is gapped by  a finite chemical potential $\mu$ term. By mirror symmetry, the top and bottom surfaces take the same sign for the surface  Chern number, thus the total  Chern number adds up to $C=\pm 1$. Each of the two surfaces contributes to a half of the thermal Hall coefficient in Eq.\ \eqref{eq:THall}.

\subsubsection{Effects of mirror symmetry breaking}
\label{sec:efield}

So far, we see that mirror symmetry ${\cal R}_z$ plays an important role in the topology of the chiral $p$-wave Weyl loop superconductor both in the bulk and in a slab geometry. In this section, we investigate the effect of a mirror symmetry breaking field $E$ in \eqref{eq:bdg}. 

With a mirror-breaking $E$ field (\ref{eq:Efield}), mirror Chern number is ill-defined, and thus the mirror-symmetry protected $xz$ and $yz$ surface states become gapped. We compute the effective low-energy Hamiltonian of those surface states in Appendix \ref{app:surface}. For $xz$ surface, the result is 
\begin{align}
h_{xz}(\kv) = \Delta k_x \eta_z - 2t_2 k_z \eta_y - E \eta_x,
\end{align}
where $\eta_{x,y,z}$ are Pauli matrices in the basis spanned by two spinors $\Psi$ with $\sigma_1\tau_x \Psi=\Psi$. 
For the $xy$ surface states that are already gapped, an $E$ field simply shift their energies. We show in Appendix \ref{app:surface} that the effective low-energy Hamiltonian of the top/bottom surface state is given by
\begin{align}\label{eq:22}
h_{xy}(\kv) = \Delta (k_x\tau_x + k_y\tau_y) - (\mu \mp E) \tau_z.
\end{align}

\begin{figure}
\centering
\includegraphics[scale=1.3]{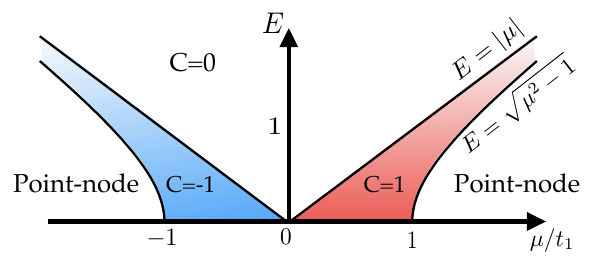}
\caption{\label{fig:phasediag2} Global phase diagram on the topology of the chiral $p$-wave superconducting state in a doped Weyl-loop semimetal  in the presence of reflection symmetry breaking electric field $E$. The topological regions with nonzero  Chern numbers are the same regions as those in which the normal state has a torus-shaped Fermi surface. This phase diagram is plotted for $t_3=0$. For $t_3\neq 0$, there will be minor modifications to the phase boundaries, while the relative locations of phases do not change. Point-node refers to gapless bulk superconductivity where the bulk gap vanishes at two points along the $z$-axis.}
\end{figure}

Moreover, we see that there is a gap closing at $|E|=|\mu|$ in either top or bottom surface, indicating an inversion of the Dirac mass. It is well-known that such a Dirac mass inversion induces a change in  Chern number by one. 
Indeed we numerically found that for $|E|<|\mu|$, the total  Chern number remains at $|C|=1$, while for $|E|>|\mu|$, the total  Chern number changes to $C=0$, and the superconducting phase becomes trivial. Note that this phase boundary at $|E|=|\mu|$ also occurs at the normal state analysis (see Fig.\ \ref{fig:phasediag}), and corresponds to the transition across which the bulk torus Fermi surface shrinks and disappears. On the other hand, as we have found, for larger $\mu$, the torus Fermi surface expands into a sphere-like topology. With a chiral $p$-wave superconducting order, it is easy to show that this phase exhibits point (Weyl) nodes at $(k_x,k_y)=0${, and the Chern number is no longer well-defined}. This is similar to the $A$-phase of 3-He. We find that as long as the normal-state Fermi surface remains torus-shaped, the  Chern number $|C|=1$ is robust and the corresponding quasi-2D phase is topological. In this sense, the topology of the quasi-2D system is a property of the torus-shaped Fermi surface. While we here have only demonstrated this point for a constant $E$ field, we anticipate that this result holds for other mirror breaking {effects}. We shall see another example of mirror breaking term in the next section.
We summarize our findings as a global phase diagram for the superconducting state in Fig.~\ref{fig:phasediag2}.

Without mirror symmetry, the $\theta=\pi$ quantization condition no longer holds. As shown in {Fig.~\ref{fig:theta}, we compute the gradual suppression of $\theta$ as $E$ is increased.} 
 Finally, we get $\theta\to 0$ as $E\to \infty$ and we reach the atomic limit where all orbitals are frozen at lattice sites. Even though $
\theta$ is no longer quantized, it still leads to a surface thermal Hall effect. 
 {As we showed in \eqref{modulo}, the surface  Chern number is defined modulo an integer and non-zero $\theta$ shows up as the anomalous fractional part. In our setup, with a positive $\mu$, to match with the mirror symmetric case with $\theta=\pi$, we have
\begin{align}
C_d=\frac{\theta}{2\pi}, ~~~C_u=1-\frac{\theta}{2\pi},
\end{align}
for $E<\mu$
and
\begin{align}
C_d=\frac{\theta}{2\pi},~~~C_u=-\frac{\theta}{2\pi},
\end{align}
for $E>\mu $ after the surface topological phase transition occurs. We have checked this fact numerically as shown in Fig.~\ref{fig:Efield}(a).}

\begin{figure}
\centering
\includegraphics[scale=.44]{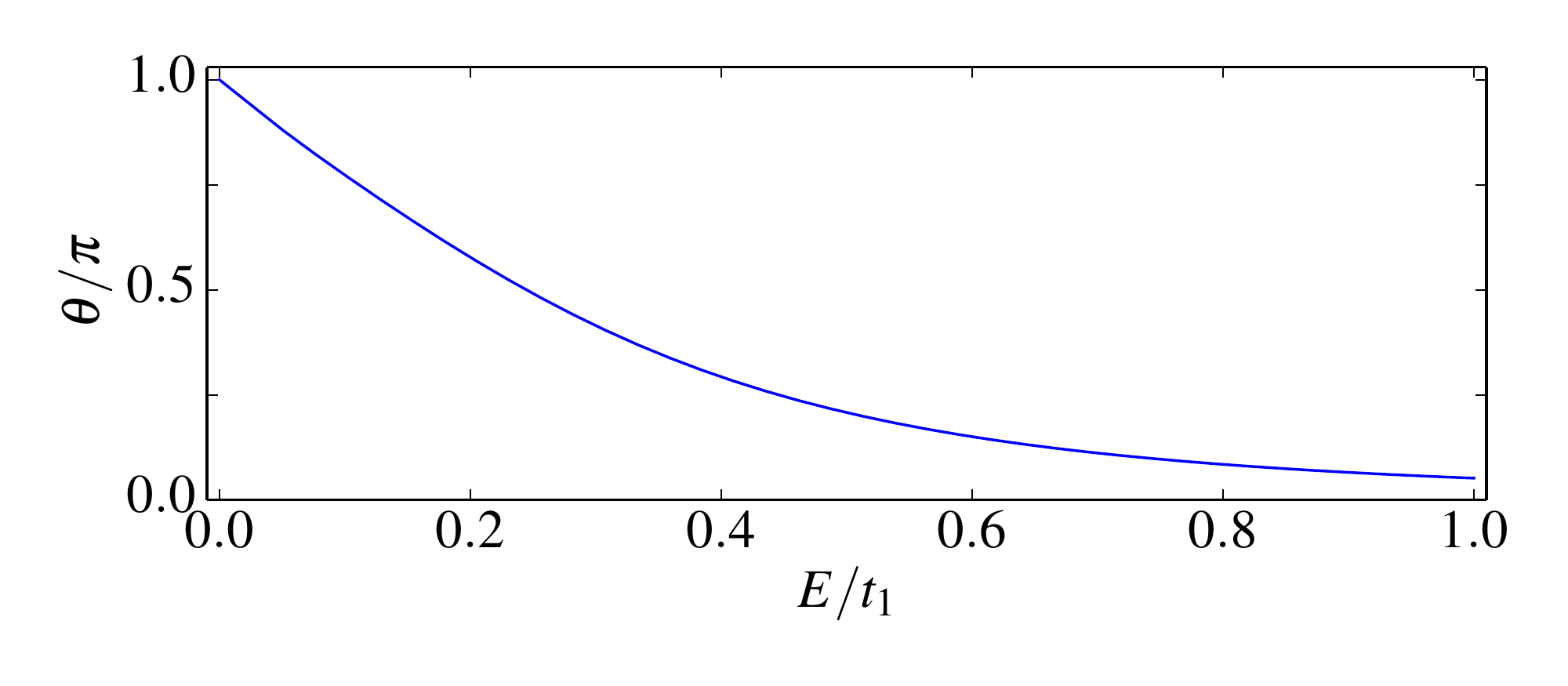}
\caption{\label{fig:theta} The axion angle as a function of electric field $E$ for the four band model (\ref{eq:bdg}) computed by Eq.~(\ref{eq:theta}). Note that $\theta$ is independent of $\mu$ and we have same curve for different values of $\mu$ as long as $|\mu|<t_1$. Here, we set $t_3=0.5$. }
\end{figure}

\begin{figure}
\centering
\includegraphics[scale=1.03]{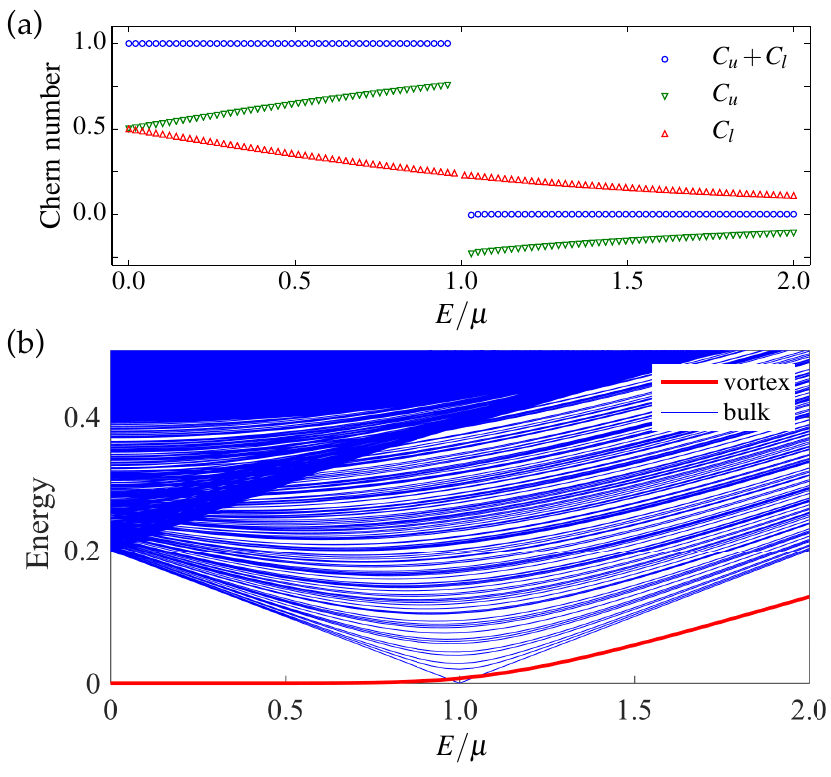}
\caption{\label{fig:Efield} (a) The evolution  of (a) the total Chern number and surface Chern numbers and (b) the energy spectrum as a function of electric field $E$ {with a fixed $\mu$}. In panel (b), blue curves show the spectrum of the Hamiltonian (\ref{eq:bdg}) for a slab geometry and the red curve indicates the nearest to zero eigenvalue in the presence of a vortex line. The system closes the gap and undergoes a topological quantum phase transition at $E/\mu=1$. Beyond that, the total Chern number vanishes and the Majorana zero-mode is lifted. 
In both panels, we set $t_3=0.5$ and $\mu=0.2$. In (b), in order to find the Majorana-zero mode in the presence of a vortex line, sparse diagonalization was done for a system of $12^3$ sites. Furthermore, the energy spectrum in the slab geometry was computed for an $80\times 80 \times 20$ system. 
}
\end{figure}

\subsubsection{Vortex core Majorana zero modes}

As a result of the Chern (winding) number $C=1$ in the quasi-2D geometry, one expects a \emph{single} Majorana mode bound to the core of a vortex line penetrating the slab in the $z$-direction. Note that this is in contrast with the well-known Fu-Kane superconductor~\cite{fu-kane-2008}, which is a $s$-wave proximitized  topological insulator~\cite{hosur-prl}. In the latter case there are \emph{two} Majorana zero modes each localized at one end of the vortex line, and can be equivalently viewed as a quasi-1D topological superconductor in class D.

We numerically confirm the existence of a single Majorana vortex-core bound state when a magnetic flux vortex of $h/2e$ is inserted through the sample along the $z$-direction for different cases in the regime $C=1$. The {energy spectrum for the bulk states and for the lowest energy vortex-line} bound state as a function of $E$ are shown in Fig.~\ref{fig:Efield}(b). We  further provide explicit examples in which we show the Fermi surface and Majorana mode wave function for a few different cases in Fig.\ \ref{fig:Majorana}.

In particular, with mirror symmetry ($E=0$), we examined two different cases, with and without a $t_3$ term in Eq.\ \eqref{eq:bdg}, which adds a quadratic dispersion, and, in our case, a ``surface Fermi contour", to the normal state drumhead surface bands. Recalling that a 2D $p_x+ip_y$ superconductor is topological (trivial) with (without) a normal state Fermi surface, it would have been tempting to claim that for our case when $t_3=0$ there is no Majorana mode, and when $t_3$ is tuned to give rise to a surface Fermi contour there are two Majorana modes, each localized at a surface. However, with the understanding of the  Chern number $C=1$, we now know that for both cases there should only be one vortex-core Majorana mode. {This is indeed so, but} interestingly, cases with and without a surface Fermi contour exhibit different profiles of the Majorana wave function. Without a surface Fermi contour, the Majorana wave function is smeared throughout the vortex line [see Fig.\ \ref{fig:Majorana}(a)], while with a surface  Fermi surface the Majorana wave function is more concentrated near the two ends of the vortex line [see Fig.\ \ref{fig:Majorana}(b)]. It is important to note that for the case with the surface Fermi contour the ``bulk" component for the vortex-line Majorana bound state does not vanish, or decay exponentially -- as in that scenario the two wave packets at the ends of the vortex line would decouple and represent two separate Majorana modes. Instead the wave function stays as a constant through the bulk part of the vortex line. Thus for both cases even though the layer-resolved  Chern numbers are localized near the top and bottom surfaces, the wave function of the Majorana mode always have a non-vanishing bulk component.  This fact indicates that {the bound-state spectrum in a vortex line of infinite length} is gapless. 
This is consistent with our earlier finding that the side surfaces host gapless Majorana cones [see Eq.~(\ref{eq:sidesurface})], {since a vortex line can be viewed as a cylindrical ``inner surface" of the sample with its radius shrunken to zero.} {The expected form of chiral boundary Majorana modes is shown in Fig.~\ref{fig:wrapping}(a). A typical density profile at zero electric field is plotted in Fig.~\ref{fig:wrapping}(c).}

\begin{figure}
\centering
\includegraphics[scale=.65]{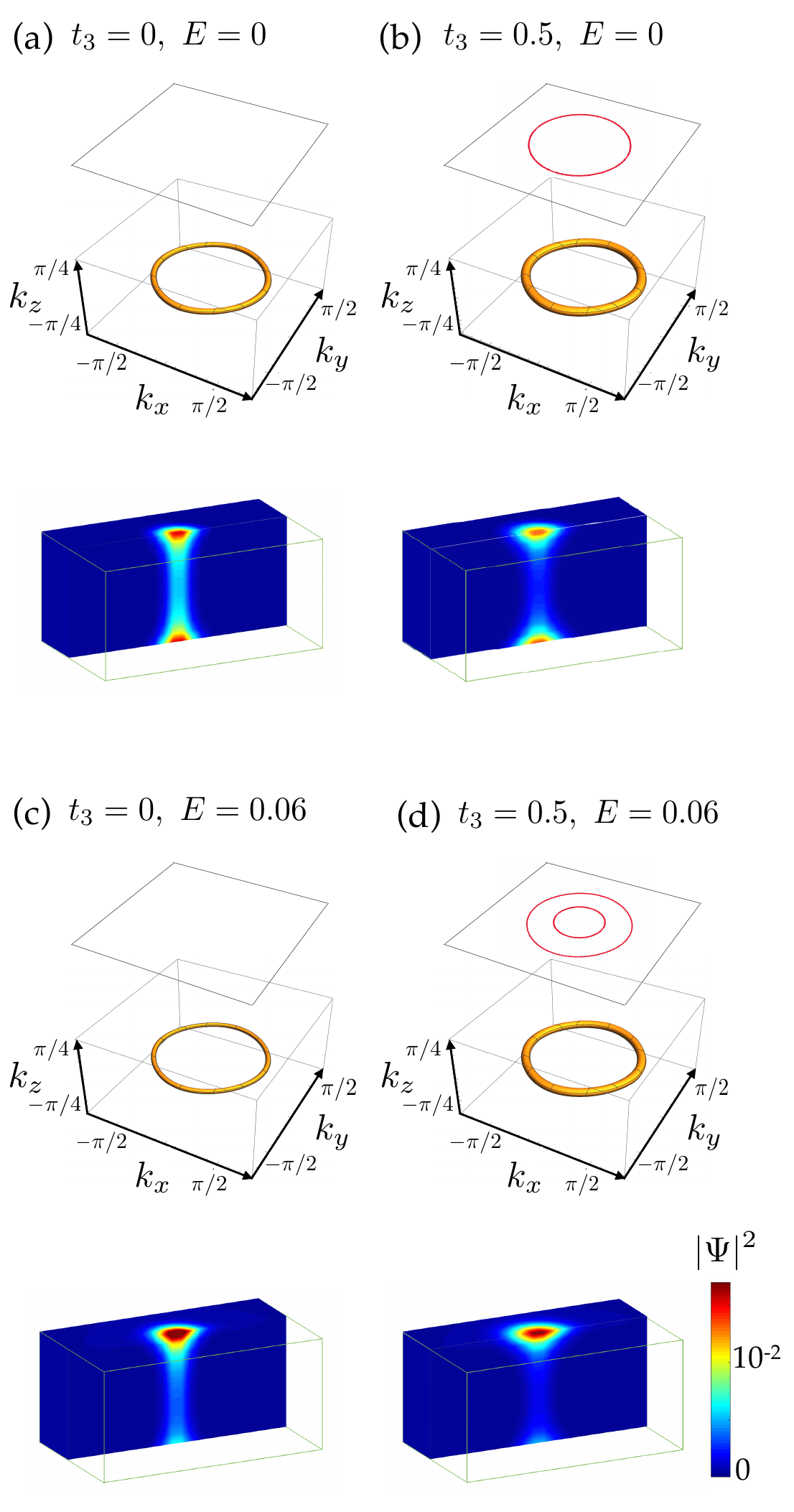}
\caption{\label{fig:Majorana}  Upper panels of (a)-(d): Bulk Fermi surface and surface Fermi contour drawn in 3D and 2D Brillouin zones (BZ). Blank 2D BZ indicates no surface Fermi contour. 
 Lower panels of (a)-(d): Three-dimensional density profile of Majorana zero modes bound to a $h/2e$ flux vortex inserted into the superconductor at the corresponding $t_3$ and $E$. Here, we set $\mu=0.1$ and $\Delta=0.4$. The system size is $12^3$.}
\end{figure}

If mirror symmetry $\mathcal{R}_z$ is broken by an $E$ field, then in the topological phase the Majorana mode is expected to be localized at a given end. We confirm this numerically, as shown in Fig.\ \ref{fig:Majorana}(c,d). Unlike in Fig.\ \ref{fig:Majorana}(a), in these cases the ``bulk" contribution to the Majorana mode does decay exponentially. To confirm the distinction between the cases with and without an $E$ field on a more quantitative level, we compute the IPR of the Majorana wave function (\ref{eq:IPR}) as shown in Fig.~\ref{fig:localization}(a). We see that in the limit of layer number $L_z\to \infty$ the $\mathrm{IPR}_\Psi$ vanishes ($|\Psi|^2\propto 1/L_z$ which leads to IPR$_\Psi \propto 1/L_z$) for the case with $E=0$ and tends to a constant for $E\neq 0$, thus the wave function for the latter case is localized. As a comparison, we verify for both cases that the distribution of the layer-resolved  Chern number $C_z$ is ``localized" by computing $\mathrm{IPR}_C$; as shown in Fig.\ \ref{fig:localization}(b) $\mathrm{IPR}_C$ tends to a {nonzero} constant as $L_z\to \infty$ with and without electric field.

\begin{figure}
\centering
\includegraphics[scale=.43]{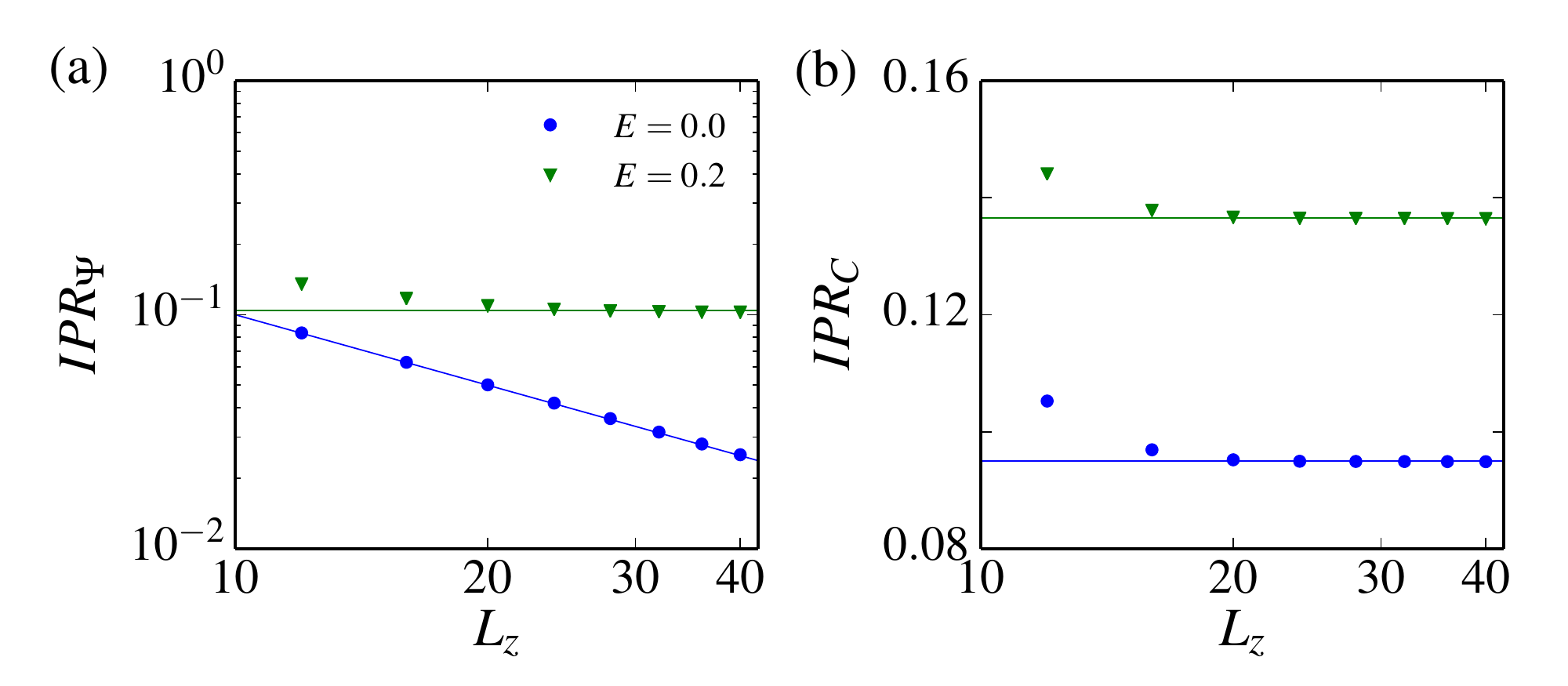}
\caption{\label{fig:localization}
  The effect of electric field on the localization as characterized by the inverse participation ratio for (a) vortex Majorana bound state and (b) layer-resolved  Chern number. The solid lines are guides for eyes and to emphasize the large $L_z$ behavior. For $E=0.2$, both Majorana bound-state and layer-resolved  Chern number are localized and IPR stays constant as the system size increases. For $E=0.0$, Majorana bound-state is delocalized IPR$_\Psi \propto 1/L_z$ (blue solid line in (a)) whereas the layer-resolved  Chern number is localized. Here, we set $t_3=0.5$, $\mu=0.5$, and $\Delta=0.4$.}
\end{figure}

The localization of the Majorana wave function in the presence of an $E$ field can also be understood via simple calculations. We sketch the key arguments and steps here, and provide the details in Appendix \ref{sec:Majorana}.
The surface component of the Majorana bound state can be obtained from \eqref{eq:22} by steps similar to the standard derivation for a 2D $p_x+ip_y$ superconductor. For $0<E<\mu$, the resulting wave function for both surfaces is
\be \label{eq:gammart}
\gamma (r, \theta)\sim \exp\[-\int ^r \frac{\mu\mp E}{\Delta(r') }dr' \] \chi(\theta),
\ee 
where $(r,\theta)$ are the 2D polar coordinates, and $\chi(\theta)=(e^{i\theta}, e^{-i\theta})^T$ is a spinor in Nambu space satisfying
\be \label{eq:tautheta}
\tau_\theta \chi \equiv \(\cos 2\theta \tau_x - \sin 2\theta \tau_y\) \chi= -\chi.
\ee
One can show that there is another solution with $\tau_\theta \chi =\chi$ has a radial part that diverges at large $r$, and hence must be discarded.
Still, the form of the wave function in (\ref{eq:gammart}, \ref{eq:tautheta}) is not normalizable near $r\to 0$. 

To resolve this issue one needs to go to large $(k_x, k_y)$, which can only be provided by bulk states localized inside the vortex line. Again drawing analogy with the 2D $p_x+ip_y$ superconductor, the $k_z$-dispersion of the vortex-line states can be analytically obtained as
\be
h_{\rm vl}(k_z)= 2\sin k_z  \tau_z \sigma_2+E\tau_z\sigma_3,
\ee
while the Nambu spinor part $\chi_{\rm vl}$ of the vortex-line bound state wave functions should satisfy 
\be\label{eq30}
\tau_\theta \sigma_1 \chi_{\rm vl} = -\chi_{\rm vl},
\ee
similar to \eqref{eq:tautheta}.
The effective low-energy Hamiltonian $h_{\rm vl}(k_z)$ has zero energy solutions that exponentially decay in $\pm z$ direction. The state localized near bottom surface satisfies
\begin{align}
\label{eq31}
\sigma_1 \chi_{\rm vl} &= - \chi_{\rm vl},
\end{align}
while the one localized near the top surface obeys
\begin{align}
\sigma_1 \chi_{\rm vl} &=  \chi_{\rm vl}.
\end{align}
Combining Eqs.\ (\ref{eq30}, \ref{eq31}), the bulk vortex-line wave function with
\begin{align}
\tau_\theta \chi_{\rm vl} &=  \chi_{\rm vl},
\end{align}
is {peaked} at the bottom surface, but the other one with
\begin{align}
\tau_\theta \chi_{\rm vl} &= - \chi_{\rm vl},
\end{align}
is {peaked} at the  top surface.

We can now connect the bulk part of the vortex line wave function with the surface part. While Eq.\ \eqref{eq:tautheta} is satisfied for both surfaces, from Eq.\ \eqref{eq30} we see that only near the \emph{top} surface the surface part and the bulk vortex-line part of the wave function can {smoothly} connect to each other. Therefore the Majorana wave function is localized at the top end of the vortex line. It is straightforward to verify that if one reverse the sign of either $\mu$ or $E$, the localization would be at the bottom surface. Thus, a chiral $p$-wave Weyl loop superconductor offers an interesting possibility of controlling the Majorana zero modes by a small mirror symmetry breaking field $E$.

Finally, we note that 
{following an analogous procedure}
as above, one obtains chiral propagating modes localized at {mirror symmetry breaking (top and bottom)} \emph{hinges} of a cubic sample~\cite{Hughes1, Hughes2, Fang, Neupert, Luka, neupert_new} when the reflection symmetry is broken by a small $E$ field (see Fig.~\ref{fig:wrapping}(b) and (d)). 
The chiral modes can be interpreted as domain-wall states formed between the gapped boundary surfaces with opposite {(layer-resolved)} Chern numbers {$C=\pm 1/2$}.
{These hinge modes have been recently analyzed in the context of the so-called ``higher-order topological insulators and superconductors"~\cite{Neupert, Luka}.}
With a trivial bulk due to broken mirror symmetry, the cube system can be viewed as a realization of the second-order topological superconductor, and in this case the second-order topological invariant is simply the Chern number $C=1$~\cite{Luka}. {Our results thus show that nodal-loop semimetals are natural platforms to realize second-order topological superconductors.}

\begin{figure}
\includegraphics[scale=0.65]{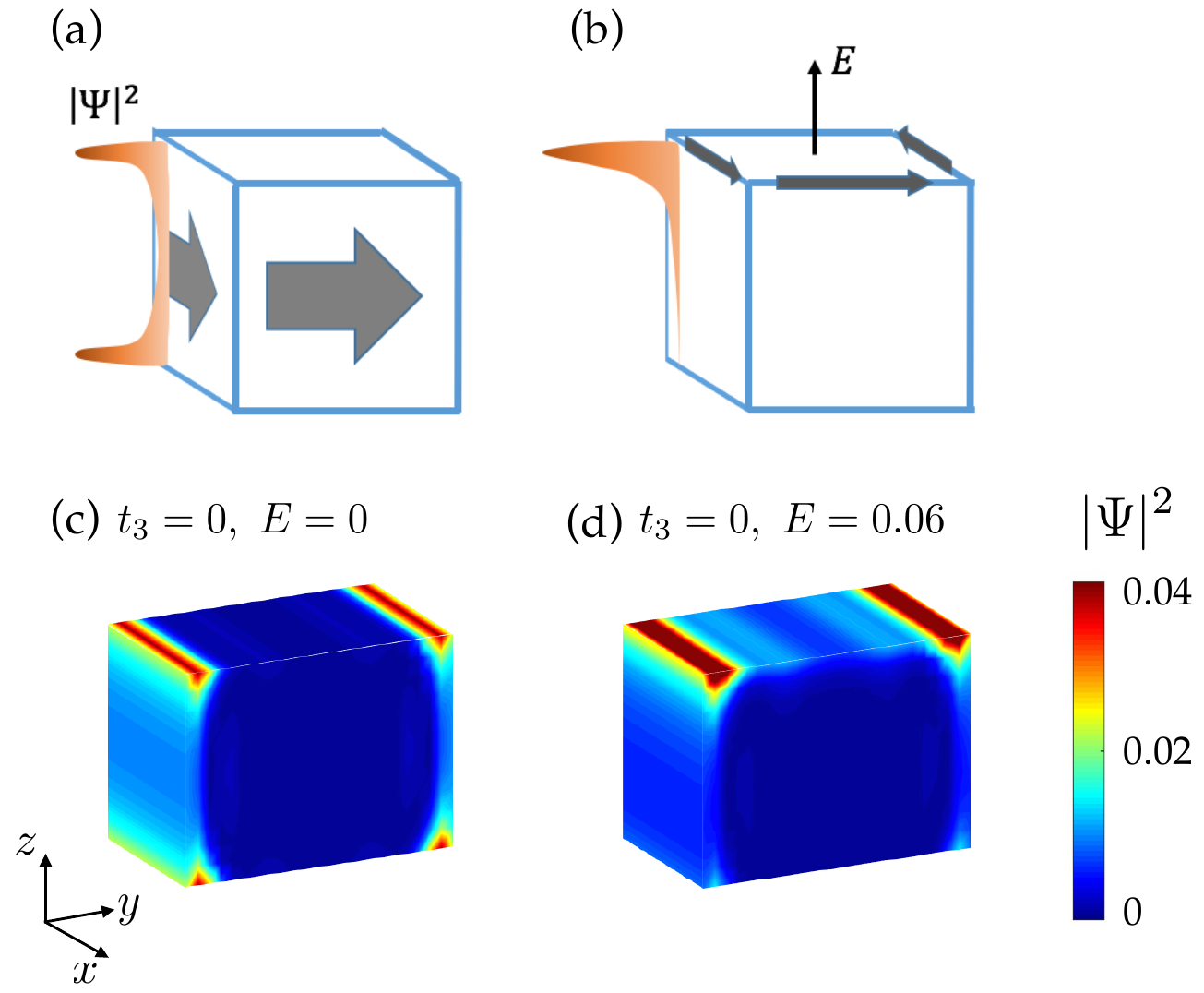}
\caption{\label{fig:wrapping} {Schematic profile of chiral Majorana boundary states  in (a) the absence and (b) presence of symmetry breaking electric field. Case (b) corresponds to hinge modes discussed in the text.
(c) and (d) show 3D density plots of the zero mode of chiral Majorana states for two scenarios of (a) and (b). In simulations, the system size is $12^3$ and other parameters are the same as   Fig.~\ref{fig:Majorana}.}
}
\end{figure}

\section{$p$-wave Dirac loop superconductors}\label{sec:dirac}

In the previous section, we have studied the intrinsic chiral $p$-wave order in a doped Weyl loop material and found that it is topological both in 3D and quasi-2D configurations. This makes the experimental realization of Weyl loop semimetal quite appealing. One candidate material for Weyl loop semimetal is TlTaSe$_2$ \cite{Bian2016} but there the Weyl loops are not centered around  $\kv =0$, and it is unclear what the leading superconducting instability is. On the other hand, to our knowledge, most nodal loop semimetal candidate materials from {first-principle} calculations and experiments exhibit a four-fold degenerate line node, i.e., a Dirac loop. 

Without spin-orbit coupling (which is indeed negligible in the candidate material CaAgP and small in CaAgAs~\cite{Yamakage2016}), a Dirac loop semimetal is nothing but two copies of Weyl loop semimetals, each copy corresponding to one spin orientation. A model Hamiltonian can be obtained by simply doubling \eqref{eq:Hcont},
\begin{align} 
{H}=\sum_\kv c_\kv^\dag \left[\frac{k_x^2+k_y^2-k_F^2}{2m}\sigma_1+ v_z k_z\sigma_2 -\mu \right]s_0 c_\kv,
\label{dirac_loop}
\end{align}
where $c_{\kv}^\dag=(c_{\kv1\uparrow}^\dag,c_{\kv2\uparrow}^\dag,c_{\kv1\downarrow}^\dag,c_{\kv2\downarrow}^\dag)$,  $s_0$ is an identity matrix in the spin space, and here $\sigma$'s denote an orbital degree of freedom. Such a Hamiltonian has the same mirror symmetry ${{\cal R}_{z}} c^\dag_{\kv} {{\cal R}_{z}}^{-1}= \sigma_1 c^\dag_{{{\cal R}_{z}}\kv}$, which protects the nodal loop.
This model also has a time-reversal symmetry, given by
\begin{align} \label{eq:tr}
{{\cal T}} c^\dag_{\kv} {{\cal T}}^{-1}= is_2 c^\dag_{-\kv}.
\end{align}

\subsection{Superconductivity in a doped Dirac loop semimetal}
Due to the extra spin degree of freedom, the structure of the superconducting order parameter is more complicated~\cite{Nandkishore2016}. Splitting the superconducting order parameter into a spin part, an orbital part, and a spatial part (in the absence of spin-orbit coupling), we can classify all possible superconducting states allowed by fermionic statistics, shown in Table \ref{tab:SC}.

 For a repulsive interaction, by the same reasoning as in the Weyl loop case~\cite{Sur-Nandkishore2016}, we expect that the leading order instability is toward an odd parity SC order, while we expect a ``conventional" $s$-wave even parity order to develop with an attractive interaction.
  Furthermore, the orbital-singlet order parameter is given by
 \be
 c^\dagger_\kv \hat{\Delta}  c^\dagger_{-\kv}=  c^\dagger_{\kv,\alpha}\Delta_{\alpha\beta}(\kv) i\sigma_2 c^\dag_{-\kv,\beta},
 \ee
 where $\alpha,\beta$ are spin indices.
 The orbital ($\sigma$) part of the order parameter is similar to that of the $s$-wave state in a doped Weyl loop superconductor. The latter state has been analyzed in Ref.~\onlinecite{Wang-Nandkishore2017}, and it was found to exhibit two line nodes at $k_z=0$. Similarly, we find that for the orbital-singlet state, there are two line nodes at $k_z=0$, aside from additional nodes given by the zeros of $\Delta_{\alpha\beta}(\kv)$. The appearance of nodal lines lowers the condensation energy; therefore, we expect the leading superconducting instability to be toward a spin-triplet, orbital-triplet, and odd parity state.
In the following, we focus on superconducting states that preserves the mirror symmetry $\mathcal{R}_z$.
  Again following the analogy in the Weyl loop case \eqref{eq:p+ip}, it is easy to show that such a fully gapped state is of the form
 \begin{align}
 c^\dagger_\kv \hat\Delta  c^\dagger_{-\kv}&=  c^\dagger_{\kv,\alpha}\Delta_{\alpha\beta}(\kv) \sigma_0 c^\dag_{-\kv,\beta}\nonumber\\
  &= c^\dagger_{\kv}{\bf d}({\kv})\cdot{\bf s}~ (is_2)~ \sigma_0 c^\dag_{-\kv},
 \end{align}
 where ${\bf d}(\kv)$ is odd in $\kv$ and does not contain odd powers of $k_z$ because of mirror symmetry ${\cal R}_z$. It is natural then to consider the following two cases. Up to a global spin rotation that equivalently rotates the orientation of $\bf d$, the first case is 
 \be \label{eq:a}
 {\bf d}({\kv})\propto (k_x + ik_y) \hat y.
 \ee
  This is an analog of the $A$-phase of $^3$He, and breaks time-reversal symmetry \eqref{eq:tr}. Following the convention, we also dub this phase for the Dirac loop superconductor the $A$-phase. 
 It is easy to see that the $A$-phase is simply two copies of the  chiral $p$-wave  Weyl loop superconductor discussed in the previous section.
 The second case, again up to a global spin rotation, corresponds to 
 \be \label{eq:b}
 {\bf d}({\kv})\propto (k_x, k_y).
 \ee
 This is an analog of the $B$-phase of $^3$He. Hence, we call this phase the $B$-phase of the Dirac loop superconductor. It is straightforward to show that the $B$-phase corresponds to a $p_x+ip_y$ order for the spin-up Weyl loop copy, and $p_x - ip_y$ order for the spin-down Weyl loop copy. Since time-reversal \eqref{eq:tr} flips spin and contains a complex conjugation, the $B$-phase is both mirror and time-reversal symmetric.

While we have provided qualitative arguments on why the spin-triplet, orbital-triplet and odd parity states, i.e., the $A$-phase and the $B$-phase are leading instabilities for a repulsive interaction, it is beyond the scope of this work to perform a detailed energetic analysis to determine the superconducting ground state. Below we will focus on the topological properties of the two phases, which can be easily elucidated based on our results of the Weyl loop superconductor.

\subsection{Topology of the $A$-phase}
The BdG Hamiltonian of  the $A$-phase of the $p$-wave Dirac loop superconductor can be expressed as
\begin{align} \label{eq:bdg2}
{\cal H} = & s_0\[(6 - t_1 -2 \cos k_x - 2 \cos k_y -2 \cos k_z) \tau_z \sigma_1 \right. \nonumber \\
&+ 2 t_2   \sin (k_z) \tau_z \sigma_2- \mu  \tau_z \sigma_0 
 \nonumber \\
 &+ t_3  (2-\cos k_x -\cos k_y)\tau_z \sigma_0 \nonumber \\
 &\left.+ \Delta  (\tau_x \sin k_x + \tau_y \sin k_y) \sigma_0\],
  \end{align}
  As mentioned above, it can be regarded simply as two identical copies of the chiral $p$-wave Weyl loop superconductor. For this reason, it is a topological crystalline superconductor in class D$+{\cal R}_+$, and here the only difference is that the strong topological invariant is
    \be
  |N_{M\Z}|=2,
  \ee
  which stabilizes \emph{two} gapless Majorana cones on each mirror invariant surface.
   In a quasi-2D slab geometry the  Chern number is $|C|=2$, which supports a thermal Hall effect with 
  \be 
\sigma_{xy}^T = \pm \frac{(\pi k_B)^2 T}{3h},
\ee
and by the same token, each surface contributes half of the thermal Hall coefficient. {Similar to the Weyl loop case, this Chern number can also be viewed as a second-order topological invariant~\cite{Luka}.}

Cases with  Chern number $|C|=2$ usually are not associated with vortex core Majorana zero modes. The reason is that the two would-be zero modes localized inside the vortex from the  Chern number can  couple  and gap out each other. We note here that this issue can be circumvented by spatially separating the two zero modes by a mirror breaking term. Consider an additional term to the BdG Hamiltonian \eqref{eq:bdg2}
\be \label{eq:bfield}
\mathcal{H}_B=Bs_3\sigma_3,
\ee
which breaks both mirror symmetry and time-reversal symmetry and can be regarded as a staggered Zeeman coupling term. Effectively, this term generates opposite $E$-field terms discussed in Eq.\ \eqref{eq:Efield} and Sec.\ \ref{sec:efield} for the two spin species, and from the analysis in Sec.\ \ref{sec:efield}, there are two Majorana zero modes bound at the vortex core, localized at the top and bottom ends respectively. This configuration of Majorana modes is similar to the case of the Fu-Kane superconductor~\cite{fu-kane-2008}, but here the topological classification is quite different. It remains to be seen if this  term  \eqref{eq:bfield} can be realized in materials, e.g., from a magnetic order.
     \begin{table}
\centering
\caption{Classification of superconducting order parameters for a doped Dirac loop semimetal in the absence of spin-orbit coupling.}
\label{tab:SC} 
\begin{ruledtabular}
 \begin{tabular}{ccc}
Spin ($s$) & Orbital ($\sigma$) & Spatial\\
  \hline
  Singlet & Singlet & Odd  \\
  Triplet & Triplet & Odd  \\
  Singlet & Triplet & Even \\
  Triplet & Singlet & Even \\
  \end{tabular}
 \end{ruledtabular}
 \end{table}
  
\subsection{Topology of the $B$-phase}
Unlike the $A$-phase, the $B$-phase preserves time-reversal symmetry. The BdG Hamiltonian is given by\begin{align} \label{eq:bdg3}
{\cal H} = & s_0\[(6 - t_1 -2 \cos k_x - 2 \cos k_y -2 \cos k_z) \tau_z \sigma_1 \right. \nonumber \\
&+ 2 t_2   \sin (k_z) \tau_z \sigma_2- \mu  \tau_z \sigma_0 
  \nonumber \\
 &\left.+ t_3  (2-\cos k_x -\cos k_y)\tau_z \sigma_0\] \nonumber \\
 &+\Delta  (s_3 \tau_x \sin k_x + \tau_y \sin k_y) \sigma_0,
  \end{align}
  and can be viewed as a $p_x+ip_y$ superconductor  for spin up and a $p_x-ip_y$ superconductor for spin down, which are related by time-reversal symmetry.
  
  From the results in Refs.\ \onlinecite{Chiu2013,Shiozaki2014}, the $B$-phase of the Dirac loop superconductor belongs to class DIII$+{\cal R}_{++}$ (the subscript $++$ means that the reflection operator $R_z=\sigma_1$ commutes with both time-reversal $is_2 {\cal K}$ and particle-hole symmetry $\tau_x {\cal K}$), and is classified by a $\mathbb{Z}_2$ invariant. 
  Similar to class D$+{\cal R}_+$, the $\mathbb{Z}_2$ invariant  for our case can be read off from the $\mathbb{Z}_2$ time-reversal topological invariants for mirror-symmetry invariant subsystem at $k_z=0$ and $k_z=\pi$ with a given mirror eigenvalue $\sigma_1=1$. Only the subsystem at $k_z=0$ turns out to be nontrivial, given by
 \begin{align}
 {\cal H}_{k_z=0}^{{{\cal R}_{z}}=+1} = & (3-\mu -2 \cos k_x - 2 \cos k_y) \tau_z  \nonumber \\
 &+ \Delta  (s_3 \tau_x \sin k_x + \tau_y  \sin k_y),
    \end{align}
    which describes a standard time-reversal invariant 2D topological superconductor. Hence, the $B$-phase of the Dirac loop superconductor is a topological crystalline superconductor in class DIII$+{\cal R}_{++}$. On the mirror-invariant surfaces, there exist gapless states protected by both mirror symmetry and time-reversal symmetry.
    
    In a slab geometry, it is easy to show that the quasi-2D system carries a nontrivial $\mathbb{Z}_2$ invariant (not to be confused with the bulk $\mathbb{Z}_2$ invariant), since the slab system can be viewed as two copies with  Chern numbers $C_\uparrow=-C_\downarrow= 1$ which are related by time-reversal symmetry.
     Due to the presence of time-reversal symmetry, there are two Majorana zero modes for a vortex line penetrating the system in $z$ direction, which are protected by time-reversal symmetry. {This $\mathbb{Z}_2$ invariant can be viewed as a second-order topological invariant~\cite{Luka}.}

\subsection{Effects of spin-orbit coupling}

So far we have assumed our Dirac loop band structure to be symmetric under mirror reflection {given by \eqref{eq:reflection}}, which protects the nodal loop. In reality, for example in candidate material CaAgAs, there exist spin-orbit coupling effects that breaks this mirror symmetry and introduces a (small) gap to the nodal loop. Such a spin-orbit coupling term turns the nodal-loop semimetal into a topological insulator at neutrality.

In terms of topology, it is clear that any mirror breaking effects invalidates the bulk topological invariants for both the $A$-phase and the $B$-phase, which are only well-defined with mirror symmetry. However, as we see for the case of Weyl loop superconductor in the presence of a mirror breaking $E$ field, the topology in the quasi-2D slab geometry may still be robust. Indeed, for the $A$-phase, as long as the spin-orbit-coupling term does not close a gap in the bulk or on the surface, the  Chern number for the slab cannot change. For the $B$-phase, the same criterion applies provided that the time-reversal symmetry is preserved.

 In this section, we examine the interplay between the superconducting order parameter $\Delta$, which is typically parametrically smaller than other energy scales of the problem with the spin-orbit coupling $\lambda$ that gaps out the nodal line (but may leave the torus Fermi surface intact). Our conclusion is that, similar to the Weyl loop case with an $E$ field, as long as the torus-shaped Fermi surface in the normal state is maintained, the quasi-2D topology remains stable even for an infinitesimal superconducting order parameter, for both $A$-phase and $B$-phase. 
 
 Let us use the superconducting $A$-phase as an example. The lattice Hamiltonian can be written as
 \begin{align}
 \mathcal{H}= &\Delta(\sin k_x \tau_x + \sin k_y \tau_y)+ 2t_2 \sin k_z \tau_z \sigma_2 \nonumber\\
 & + (6-t_1-2\cos k_x -2\cos k_y -2 \cos k_z) \tau_z \sigma_1  - \mu \tau_z \nonumber\\
& + \lambda(\sin k_x  s_1 \sigma_3 + \sin k_y s_2   \tau_z \sigma_3),
 \label{HDL}
 \end{align}
 where $\sigma$ is an orbital index, $s$ is spin, and $\tau$ is the Nambu matrices. We should note that the realistic Hamiltonian for CaAgAs is more complicated~\cite{Yamakage2016}
  but our analysis here can be straightforwardly extended to the realistic case.   
 One can see that in both Hamiltonians, both $\Delta$ and $\lambda$ gap out the nodal line for $\mu=0$, and as such they compete with each other. For $\Delta=0$, it is easy to see that this Hamiltonian actually describes a strong topological insulator.
  Since typically the superconducting gap $\Delta$ is small, a natural question is whether the topology of a quasi-2D system induced by superconductivity is spoiled by a finite spin-orbit coupling term.
 
First, it is clear that the the bulk is always gapped for an arbitrary magnitude of $\lambda$. We will take a slab geometry and look for possible gap closing on the surfaces. From \eqref{HDL} and similar calculations provided in Appendix \ref{app:surface},  we see that the surface Hamilton is given by
\begin{align}
h_{\rm surf} = \Delta( k_x \tau_x +  k_y \tau_y) - \mu \tau_z 
+ \lambda( k_x  s_1 \sigma_3 +  k_y s_2\tau_z \sigma_3 ). \nonumber
\end{align}
Without either $\lambda$ or $\Delta$, this is a gapped Dirac cone system. In the continuum limit, with both $\lambda$ or $\Delta$ it has point nodes along $k_x$ direction for $\Delta<\lambda$:
\be
E_{\rm surf}(k_x)= \pm \sqrt{\Delta^2 k_x^2 + \mu^2} \pm \lambda k_x,
\ee
and the nodal point occurs at $k_x^0 = \mu / \sqrt{\lambda^2- \Delta^2}$.
However, recall that we are dealing with surface drumhead bands whose momentum range is bounded by the size of the nodal ring $k_F$, given in the lattice model by $2\cos k_F =2-t_1$. Thus as long as $k_x^0 > k_F$, the surface bands remain gapped, and the topology of the quasi-2D system is unchanged. This corresponds to the following condition
\begin{align}
\sqrt{\mu^2+\Delta^2 k_F^2} >  \lambda k_F.
\end{align}
We see that as long as $\mu > \lambda k_F$, the gap, and hence topology, are robust against the spin-orbit coupling term, \emph{no matter} how small $\Delta$ is. A simple calculation shows that this is precisely the condition that the 
normal state torus FS remains intact in the presence of spin-orbit coupling.

Similarly the calculation can be extended to the surface states of the $B$-phase and the conclusion is the same: as long as the normal state has a torus-shaped Fermi surface, the order parameters described in Eqs.\ (\ref{eq:a}, \ref{eq:b}) give rise to topological superconducting states in a slab geometry.

\section{\label{sec:conclusions} Conclusion}
In this work, we have systematically examined the fully gapped $p$-wave superconducting states in doped nodal loop semimetals with a torus-shaped Fermi surface, for both the Weyl loop case and  the Dirac loop case. We have found that for both cases the $p$-wave states are natural realizations and perfect testbeds for topological crystalline superconductors protected by a mirror symmetry. The mirror symmetry is perpendicular to the nodal loop direction, the same one that is invoked in the normal state to stabilize the nodal loop. As a result, the mirrror preserving surfaces host gapless states characterized by strong topological invariants. 

Aside from the topology as 3D systems, quasi-2D systems in a slab geometry also carry nontrivial topology. For the Weyl loop superconductor and the $A$-phase of the Dirac loop superconductor,  the topology is characterized by a  Chern number, which leads to a quantized thermal Hall effect and possible vortex core Majorana zero modes. For the $B$-phase of the Dirac loop superconductor, time-reversal symmetry is preserved and the topology is characterized by a $\mathbb{Z}_2$ invariant. 
These Chern numbers and $\mathbb{Z}_2$ invariant have been recently discussed as second-order topological invariants~\cite{Luka} of the system  in a cube geometry, which induces hinge modes if the mirror symmetry is broken. In this sense the nodal loop superconductors also provide potential realizations of second-order topological superconductors~\cite{Hughes1, Hughes2, Fang, Neupert, Luka,neupert_new}.

With regard to experiments, we note that in the candidate Dirac loop material
CaAgAs the mirror symmetry $\mathcal{R}_z$ is only approximate and is broken by
a spin-orbit coupling term. The broken mirror symmetry thus spoils the bulk
topological invariants for both the $A$ and $B$ states. However, as we show in this paper,
the \emph{quasi-2D} topological invariants in both cases are more robust,  very similar to the Weyl loop case in the presence of a mirror breaking field.
By a simple model calculation, we verify that even though the nodal loop is gapped by the spin-orbit coupling, as long as the normal state Fermi surface maintains its torus shape, the quasi-2D topology of the superconducting state is stable 
even for an \emph{infinitesimal} superconducting gap compared to the spin-orbit coupling strength. 
Therefore, it would be quite interesting to search for exotic superconducting states in nodal-loop materials.

There are currently intensive experimental efforts to search for nodal loop materials, but the focus has been mainly on the band structure of the normal state.
 Our theoretical results suggest that the torus-shaped Fermi surface in nodal loop materials is a natural host of many unconventional superconducting phases, and can potentially open a new route toward realizing topological superconductivity.

\acknowledgements

We would like to thank Taylor Hughes, Rahul Nandkishore and Luka Trifunovic for insightful discussions.
Computational resources were provided by the Taub campus cluster at the
University of Illinois at Urbana-Champaign.
  This work was supported by
  the NSF under Grant No. DMR-1455296 (HS and SR)
and by the Gordon and Betty Moore Foundation's EPiQS Initiative through Grant No. GBMF4305 at the University of Illinois (YW).

\appendix

\section{Derivation of surface modes}
\label{app:surface}
Here, we derive the surface state wave functions and show that each surface spectrum can be written in terms of $p_x+ip_y$ superconductor. 
To find the particle-hole/pseudospin content of the surface states we need to solve the domain-wall problem in which we find the effective model at the interface between the topological superconductor and a trivial matter. 

\underline{\emph{xy-surface}}:

This is modeled by a spatial-varying mass term $t_1(z)$. We shall solve for the in-gap states near the  bottom surface, i.e., topological superconductor-vacuum interface at $z=0$ where the upper-half ($z>0$) is filled with topological superconductor $t_1(z)>0$ and the lower-half ($z<0$) is filled with trivial superconductor $t_1(z)<0$ (that is topologically equivalent to vacuum). A similar solution can be found for the top surface.
Let us start by finding the zero-mode states on the topological side. The Schr\"odinger equation for the zero-momentum state $(k_x,k_y)=(0,0)$ is given by
\begin{align}
\left[ 2 t_2 \sigma_2 (-i{\partial_z}) -t_1 \sigma_1  \right] \tau_z \ket{\Psi} =0
\end{align}
the solution of which can be written as  $|\Psi\rangle=|\psi_s\rangle\otimes|\psi_p\rangle$ where $|\psi_s\rangle$ is in the pseudospin space and $|\psi_p\rangle$ is in the Nambu space. Hence, there are two solutions
\begin{align}
|\Psi_1\rangle&=e^{-t_1z/2t_2}\ket{\uparrow}\otimes \left(
\begin{array}{c}
1 \\ 0
\end{array} \right) , \\
|\Psi_2\rangle&=e^{-t_1 z/2t_2}\ket{\uparrow}\otimes \left(
\begin{array}{c}
0 \\ 1
\end{array} \right),
\end{align}
The low energy Hamiltonian in terms of in-plane momenta ${\bf k}_{\parallel}=(k_x,k_y)$ becomes, 
\begin{align}
h({\bf k}) = \Delta (k_x \tau_x + k_y \tau_y) - \mu \tau_z,
\end{align}
in the basis spanned by $\{\ket{\Psi_1},\ket{\Psi_2}\}$. 

In the presence of a mirror-breaking  $ -E\sigma_3\tau_z$ term, note that the subspace of $\{\ket{\Psi_1},\ket{\Psi_2}\}$ is closed under matrix operator $\sigma_3\tau_z$, which projects to $\tau_z$. Therefore, with a mirror-breaking $E$ field, the $xy$ bottom surface states are described by
 \begin{align}
h({\bf k}) = \Delta (k_x \tau_x + k_y \tau_y) - (\mu+E) \tau_z.
\end{align}
By the same token, one can show that the effective low-energy Hamiltonian for $xy$ top surface states are given by
 \begin{align}
h({\bf k}) = \Delta (k_x \tau_x + k_y \tau_y) - (\mu-E) \tau_z.
\end{align}

\underline{\emph{yz-surface}}:
We are to solve the domain-wall equation
\begin{align} \label{eq:dwyz}
\left[ \Delta \sigma_0\tau_x (-i{\partial_x}) -t_1 \sigma_1\tau_z + \mu \sigma_0 \tau_z  \right] \ket{\Psi} =0
\end{align}
The  two solutions are
\begin{align}
|\Psi_1\rangle&=\frac{e^{-(t_1+\mu) x/\Delta}}{\sqrt{2}}\ket{\leftarrow}\otimes \left(
\begin{array}{c}
1 \\ i
\end{array} \right), \\
|\Psi_2\rangle&=\frac{e^{-(t_1-\mu) x/\Delta}}{\sqrt{2}}\ket{\rightarrow}\otimes \left(
\begin{array}{c}
1 \\ -i
\end{array} \right) .
\end{align}

The low energy Hamiltonian in terms of in-plane momenta ${\bf k}_{\parallel}=(k_y,k_z)$ becomes, 
\begin{align}
h({\bf k}) = \Delta k_y \eta_z + 2t_2  k_z \eta_y,
\end{align}
where $\eta_i$ are Pauli matrices defined in the basis spanned by $\{\ket{\Psi_1},\ket{\Psi_2}\}$ which satisfy $\sigma_1 \tau_y  \ket{\Psi_i}=-\ket{\Psi_{i}}$.  

In the presence of a mirror-breaking  $- E\sigma_3\tau_z$ term, note that the subspace of $\{\ket{\Psi_1},\ket{\Psi_2}\}$ is closed under matrix operator $\sigma_3\tau_z$, which projects to $\eta_x$. Therefore, with a mirror-breaking $E$ field, the $yz$ surface states are described by
 \begin{align}
h({\bf k}) = \Delta k_y \eta_z + 2t_2  k_z \eta_y - E \eta_x,
\end{align}
which is gapped.

\underline{\emph{xz-surface}}:
We are to solve the domain-wall equation
\begin{align}
\left[ \Delta \sigma_0\tau_y (-i{\partial_y}) -t_1 \sigma_1\tau_z + \mu\sigma_0\tau_z  \right] \ket{\Psi} =0
\end{align}
The  two solutions are
\begin{align}
|\Psi_1\rangle&=\frac{e^{-(t_1+\mu)y/\Delta}}{\sqrt{2}}\ket{\rightarrow}\otimes \left(
\begin{array}{c}
1 \\ 1
\end{array} \right) , \\
|\Psi_2\rangle&=\frac{e^{-(t_1-\mu) y/\Delta}}{\sqrt{2}}\ket{\leftarrow}\otimes \left(
\begin{array}{c}
1 \\ -1
\end{array} \right).
\end{align}

The low energy Hamiltonian in terms of in-plane momenta ${\bf k}_{\parallel}=(k_y,k_z)$ becomes, 
\begin{align}
h({\bf k}) = \Delta k_x \eta_z - 2 t_2 k_z  \eta_y,
\end{align}
where $\eta_i$ are Pauli matrices defined in the basis spanned by $\{\ket{\Psi_1},\ket{\Psi_2}\}$ which satisfy $\sigma_1 \tau_x  \ket{\Psi_i}=\ket{\Psi_{i}}$.  
Similar to the previous analysis, with a mirror-breaking $E$ field, the $xz$ surface states are described by
 \begin{align}
h({\bf k}) = \Delta k_x \eta_z - 2 t_2 k_z  \eta_y - E \eta_x,
\end{align}
which is gapped.


\section{Calculation details on vortex line Majorana bound state}
\label{sec:Majorana}

In this Appendix we provide some details on the computation of the Majorana bound state in the chiral $p$-wave Weyl loop superconductor. We show how to obtain the surface part and bulk part of the vortex line mode. The calculation is similar to that for Majorana zero modes in 2D $p_x+ip_y$ superconductors.

For the surface part, one subtlety is that with a vortex configuration, the $p$-wave superconducting order parameter depends both on coordinate and momentum, which do not commute. The standard way to circumvent this complication~\cite{ivanov,Nishida2010,kvorning} is to take the anti-commutator of the coordinate-dependent part and the momentum-dependent part in the first-quantized BdG Hamiltonian. Here we show this issue is automatically taken care of if one treats the problem with the second-quantized Hamiltonian, which is given by $H=\int d^2r\,\mathcal{H}(\bf r)$, and from \eqref{eq:22}
\begin{widetext}
\begin{align}
\mathcal{H}(\r)=&\frac{1}{2}\[-(\mu\mp E)\psi^\dagger(\r) \psi(\r) + (\mu\mp E) \psi(\r)  \psi^\dagger(\r) + \Delta (r) e^{i\theta} \psi^\dagger (\r)(\del_x+i\del_y) \psi^\dagger(\r) - \Delta (r) e^{-i\theta} \psi(\r) (\del_x -i\del_y) \psi(\r)   \]\nonumber\\
=&\frac{1}{2}\[-(\mu\mp E)\psi^\dagger(\r)\psi(\r) + (\mu\mp E) \psi(\r)  \psi^\dagger(\r) \right. \nonumber\\
&\left.+ \Delta (r) e^{2i\theta} \psi^\dagger (\r)\(\frac{\del}{\del r}+\frac{i}{r}\frac\del{\del\theta}\) \psi^\dagger(\r) - \Delta (r) e^{-2i\theta} \psi(\r) \(\frac{\del}{\del r}-\frac{i}{r}\frac\del{\del\theta}\) \psi(\r)   \],
\end{align}
where $\psi$ is the fermion field, $r,\theta$ are polar coordinates, and in the second line we have used the relation $\del_x+i\del_y\equiv e^{i\theta} (\del_r + i\del_\theta/r)$. The term with $\mu\mp E$ corresponds to top (bottom) surface, and we assume that $0<E<\mu$ here.

 In the second-quantized language, the Majorana zero mode condition is
\be
[H,\gamma] = 0,
\ee
where
\begin{align}
\gamma=\int d^2r\  g(r)\[ f_1(\theta) \psi^\dagger (\r) + f_2 (\theta) \psi(\r)\].
\end{align}
Then we obtain
\begin{align}
-\mu g(r) f_1(\theta) + \Delta(r) e^{2i\theta} \( \frac\del{\del r} - \frac{1}{2r} +\frac{1}{2}\frac{\del \log \Delta(r)}{\del r} + \frac{i}{r}\frac{\del}{\del\theta}\) g(r) f_2(\theta)=&0, \nonumber\\
\mu g(r) f_2(\theta) - \Delta(r) e^{-2i\theta} \( \frac\del{\del r} - \frac{1}{2r} +\frac{1}{2}\frac{\del \log \Delta(r)}{\del r} - \frac{i}{r}\frac{\del}{\del\theta}\) g(r) f_1(\theta)=&0.
\label{bdg}
\end{align}
\end{widetext}
The solution of this set of equations is given by
\begin{align}
f_1(&\theta)=e^{i\theta},~~~f_2(\theta)=-e^{-i\theta},\nonumber\\
g(r)=&\frac1{\sqrt{r\Delta{(r)}}}\exp\[-\int^r \frac{\mu\mp E}{\Delta(r')} dr'\].
\label{wf}
\end{align}
For positive $\mu$ and $\Delta$ we can verify that the wave function given by this form is concentrated near $r=0$. 
Also, note that for a negative $\mu$, the angular part of the solution becomes
\begin{align}
f_1(&\theta)=e^{i\theta},~~~f_2(\theta)=e^{-i\theta}.
\end{align}
In other words, the spinor $\chi\equiv(f_1, f_2)^T$ in the BdG space is an eigenstate with 
\be
\tau_\theta \chi \equiv (\cos 2\theta \tau_x - \sin 2\theta \tau_y) \chi = \mp \chi,
\label{eq8}
\ee
where the $\mp$ corresponds to positive and negative $\mu$, respectively.

For the bulk states that are localized near the vortex line,  let's first split the superconducting Hamiltonian into two parts $\mathcal{H}=\mathcal{H}_{xy}+\mathcal{H}_{z}$, where
\begin{align}
\mathcal{H}_{xy}=&(3-2\cos k_x-2\cos k_y)\tau_z \sigma_1  - \mu\tau_z \nonumber \\
&+\Delta(\sin k_x\tau_x + \sin k_y \tau_y) \nonumber\\
\mathcal{H}_z =& 2\sin k_z  \tau_z \sigma_2+E\tau_z\sigma_3,
\end{align}
where we have included the mirror-breaking $E$ field into $\mathcal{H}_z$.
 We observe that $\mathcal{H}_{xy}$ is nothing but two decoupled copies of 2D $p_x + i p_y$ superconductor that we analyzed in the previous section. The two copies correspond to $\sigma_1=\pm 1$,
and importantly, one of them ($\sigma_1=1$) has an electron-like FS and the other a hole-like one.
Note also that the Fermi energies for the two copies are \emph{not} $\mu$, but rather the distance from chemical ponential to band bottom (for the elctron-like band) or top (for the hole-like band), which are of opposite signs.

At the vortex core, $\mathcal{H}_{xy}$ has two zero modes. By analogy with \eqref{eq8}, we have for the Nambu spinor part of the bulk vortex line mode, $\chi_{\rm vl}$
\be
\tau_\theta\sigma_1 \chi_{\rm vl}=-\chi_{\rm vl}.
\label{eq10}
\ee
It is easy to check that this condition commutes with $\mathcal{H}_z$. Thus we obtain that the vortex line state are described by the Hamiltonian
\be
{h}_{\rm vl}= 2\sin k_z  \tau_z\sigma_2 +E\tau_z\sigma_3.
\label{eq11}
\ee
This low-energy Hamiltonian also has zero energy solutions. Given $E>0$, its wave function is exponentially decaying in $\pm z$ direction, corresponding to \be
\sigma_1 \chi_{\rm vl}=\pm \chi_{\rm vl}.
\label{eq12}
\ee
We have used Eqs.\ (\ref{wf}, \ref{eq8}, \ref{eq10}, \ref{eq11}, \ref{eq12}) in the main text.

\bibliography{refs}

\begin{thebibliography}{62}%
\makeatletter
\providecommand \@ifxundefined [1]{%
 \@ifx{#1\undefined}
}%
\providecommand \@ifnum [1]{%
 \ifnum #1\expandafter \@firstoftwo
 \else \expandafter \@secondoftwo
 \fi
}%
\providecommand \@ifx [1]{%
 \ifx #1\expandafter \@firstoftwo
 \else \expandafter \@secondoftwo
 \fi
}%
\providecommand \natexlab [1]{#1}%
\providecommand \enquote  [1]{``#1''}%
\providecommand \bibnamefont  [1]{#1}%
\providecommand \bibfnamefont [1]{#1}%
\providecommand \citenamefont [1]{#1}%
\providecommand \href@noop [0]{\@secondoftwo}%
\providecommand \href [0]{\begingroup \@sanitize@url \@href}%
\providecommand \@href[1]{\@@startlink{#1}\@@href}%
\providecommand \@@href[1]{\endgroup#1\@@endlink}%
\providecommand \@sanitize@url [0]{\catcode `\\12\catcode `\$12\catcode
  `\&12\catcode `\#12\catcode `\^12\catcode `\_12\catcode `\%12\relax}%
\providecommand \@@startlink[1]{}%
\providecommand \@@endlink[0]{}%
\providecommand \url  [0]{\begingroup\@sanitize@url \@url }%
\providecommand \@url [1]{\endgroup\@href {#1}{\urlprefix }}%
\providecommand \urlprefix  [0]{URL }%
\providecommand \Eprint [0]{\href }%
\providecommand \doibase [0]{http://dx.doi.org/}%
\providecommand \selectlanguage [0]{\@gobble}%
\providecommand \bibinfo  [0]{\@secondoftwo}%
\providecommand \bibfield  [0]{\@secondoftwo}%
\providecommand \translation [1]{[#1]}%
\providecommand \BibitemOpen [0]{}%
\providecommand \bibitemStop [0]{}%
\providecommand \bibitemNoStop [0]{.\EOS\space}%
\providecommand \EOS [0]{\spacefactor3000\relax}%
\providecommand \BibitemShut  [1]{\csname bibitem#1\endcsname}%
\let\auto@bib@innerbib\@empty
\bibitem [{\citenamefont {{Armitage}}\ \emph {et~al.}(2017)\citenamefont
  {{Armitage}}, \citenamefont {{Mele}},\ and\ \citenamefont
  {{Vishwanath}}}]{armitage-review}%
  \BibitemOpen
  \bibfield  {author} {\bibinfo {author} {\bibfnamefont {N.~P.}\ \bibnamefont
  {{Armitage}}}, \bibinfo {author} {\bibfnamefont {E.~J.}\ \bibnamefont
  {{Mele}}}, \ and\ \bibinfo {author} {\bibfnamefont {A.}~\bibnamefont
  {{Vishwanath}}},\ }\href@noop {} {\bibfield  {journal} {\bibinfo  {journal}
  {ArXiv e-prints}\ } (\bibinfo {year} {2017})},\ \Eprint
  {http://arxiv.org/abs/1705.01111} {arXiv:1705.01111 [cond-mat.str-el]}
  \BibitemShut {NoStop}%
\bibitem [{\citenamefont {Yan}\ and\ \citenamefont
  {Felser}(2017)}]{yan-review}%
  \BibitemOpen
  \bibfield  {author} {\bibinfo {author} {\bibfnamefont {B.}~\bibnamefont
  {Yan}}\ and\ \bibinfo {author} {\bibfnamefont {C.}~\bibnamefont {Felser}},\
  }\href {\doibase 10.1146/annurev-conmatphys-031016-025458} {\bibfield
  {journal} {\bibinfo  {journal} {Annual Review of Condensed Matter Physics}\
  }\textbf {\bibinfo {volume} {8}},\ \bibinfo {pages} {337} (\bibinfo {year}
  {2017})},\ \Eprint
  {http://arxiv.org/abs/https://doi.org/10.1146/annurev-conmatphys-031016-025458}
  {https://doi.org/10.1146/annurev-conmatphys-031016-025458} \BibitemShut
  {NoStop}%
\bibitem [{\citenamefont {Carter}\ \emph {et~al.}(2012)\citenamefont {Carter},
  \citenamefont {Shankar}, \citenamefont {Zeb},\ and\ \citenamefont
  {Kee}}]{Carter2012}%
  \BibitemOpen
  \bibfield  {author} {\bibinfo {author} {\bibfnamefont {J.-M.}\ \bibnamefont
  {Carter}}, \bibinfo {author} {\bibfnamefont {V.~V.}\ \bibnamefont {Shankar}},
  \bibinfo {author} {\bibfnamefont {M.~A.}\ \bibnamefont {Zeb}}, \ and\
  \bibinfo {author} {\bibfnamefont {H.-Y.}\ \bibnamefont {Kee}},\ }\href
  {\doibase 10.1103/PhysRevB.85.115105} {\bibfield  {journal} {\bibinfo
  {journal} {Phys. Rev. B}\ }\textbf {\bibinfo {volume} {85}},\ \bibinfo
  {pages} {115105} (\bibinfo {year} {2012})}\BibitemShut {NoStop}%
\bibitem [{\citenamefont {Chen}\ \emph {et~al.}(2015)\citenamefont {Chen},
  \citenamefont {Lu},\ and\ \citenamefont {Kee}}]{Chen2015}%
  \BibitemOpen
  \bibfield  {author} {\bibinfo {author} {\bibfnamefont {Y.}~\bibnamefont
  {Chen}}, \bibinfo {author} {\bibfnamefont {Y.-M.}\ \bibnamefont {Lu}}, \ and\
  \bibinfo {author} {\bibfnamefont {H.-Y.}\ \bibnamefont {Kee}},\ }\href
  {\doibase 10.1038/ncomms7593} {\bibfield  {journal} {\bibinfo  {journal}
  {Nature Communications}\ }\textbf {\bibinfo {volume} {6}},\ \bibinfo {pages}
  {6593} (\bibinfo {year} {2015})}\BibitemShut {NoStop}%
\bibitem [{\citenamefont {Schaffer}\ \emph {et~al.}(2015)\citenamefont
  {Schaffer}, \citenamefont {Lee}, \citenamefont {Lu},\ and\ \citenamefont
  {Kim}}]{Schaffer2015}%
  \BibitemOpen
  \bibfield  {author} {\bibinfo {author} {\bibfnamefont {R.}~\bibnamefont
  {Schaffer}}, \bibinfo {author} {\bibfnamefont {E.~K.-H.}\ \bibnamefont
  {Lee}}, \bibinfo {author} {\bibfnamefont {Y.-M.}\ \bibnamefont {Lu}}, \ and\
  \bibinfo {author} {\bibfnamefont {Y.~B.}\ \bibnamefont {Kim}},\ }\href
  {\doibase 10.1103/PhysRevLett.114.116803} {\bibfield  {journal} {\bibinfo
  {journal} {Phys. Rev. Lett.}\ }\textbf {\bibinfo {volume} {114}},\ \bibinfo
  {pages} {116803} (\bibinfo {year} {2015})}\BibitemShut {NoStop}%
\bibitem [{\citenamefont {Kim}\ \emph {et~al.}(2015)\citenamefont {Kim},
  \citenamefont {Wieder}, \citenamefont {Kane},\ and\ \citenamefont
  {Rappe}}]{Kim2015}%
  \BibitemOpen
  \bibfield  {author} {\bibinfo {author} {\bibfnamefont {Y.}~\bibnamefont
  {Kim}}, \bibinfo {author} {\bibfnamefont {B.~J.}\ \bibnamefont {Wieder}},
  \bibinfo {author} {\bibfnamefont {C.~L.}\ \bibnamefont {Kane}}, \ and\
  \bibinfo {author} {\bibfnamefont {A.~M.}\ \bibnamefont {Rappe}},\ }\href
  {\doibase 10.1103/PhysRevLett.115.036806} {\bibfield  {journal} {\bibinfo
  {journal} {Phys. Rev. Lett.}\ }\textbf {\bibinfo {volume} {115}},\ \bibinfo
  {pages} {036806} (\bibinfo {year} {2015})}\BibitemShut {NoStop}%
\bibitem [{\citenamefont {Mullen}\ \emph {et~al.}(2015)\citenamefont {Mullen},
  \citenamefont {Uchoa},\ and\ \citenamefont {Glatzhofer}}]{Mullen2015}%
  \BibitemOpen
  \bibfield  {author} {\bibinfo {author} {\bibfnamefont {K.}~\bibnamefont
  {Mullen}}, \bibinfo {author} {\bibfnamefont {B.}~\bibnamefont {Uchoa}}, \
  and\ \bibinfo {author} {\bibfnamefont {D.~T.}\ \bibnamefont {Glatzhofer}},\
  }\href {\doibase 10.1103/PhysRevLett.115.026403} {\bibfield  {journal}
  {\bibinfo  {journal} {Phys. Rev. Lett.}\ }\textbf {\bibinfo {volume} {115}},\
  \bibinfo {pages} {026403} (\bibinfo {year} {2015})}\BibitemShut {NoStop}%
\bibitem [{\citenamefont {Zeng}\ \emph {et~al.}(2015)\citenamefont {Zeng},
  \citenamefont {Fang}, \citenamefont {Chang}, \citenamefont {Chen},
  \citenamefont {Hsieh}, \citenamefont {Bansil}, \citenamefont {Lin},\ and\
  \citenamefont {Fu}}]{Zeng2015}%
  \BibitemOpen
  \bibfield  {author} {\bibinfo {author} {\bibfnamefont {M.}~\bibnamefont
  {Zeng}}, \bibinfo {author} {\bibfnamefont {C.}~\bibnamefont {Fang}}, \bibinfo
  {author} {\bibfnamefont {G.}~\bibnamefont {Chang}}, \bibinfo {author}
  {\bibfnamefont {Y.-A.}\ \bibnamefont {Chen}}, \bibinfo {author}
  {\bibfnamefont {T.}~\bibnamefont {Hsieh}}, \bibinfo {author} {\bibfnamefont
  {A.}~\bibnamefont {Bansil}}, \bibinfo {author} {\bibfnamefont
  {H.}~\bibnamefont {Lin}}, \ and\ \bibinfo {author} {\bibfnamefont
  {L.}~\bibnamefont {Fu}},\ }\href {http://arxiv.org/abs/1504.03492} {\bibfield
   {journal} {\bibinfo  {journal} {arXiv:1504.03492}\ } (\bibinfo {year}
  {2015})}\BibitemShut {NoStop}%
\bibitem [{\citenamefont {Weng}\ \emph {et~al.}(2015)\citenamefont {Weng},
  \citenamefont {Fang}, \citenamefont {Fang}, \citenamefont {Bernevig},\ and\
  \citenamefont {Dai}}]{Weng2015}%
  \BibitemOpen
  \bibfield  {author} {\bibinfo {author} {\bibfnamefont {H.}~\bibnamefont
  {Weng}}, \bibinfo {author} {\bibfnamefont {C.}~\bibnamefont {Fang}}, \bibinfo
  {author} {\bibfnamefont {Z.}~\bibnamefont {Fang}}, \bibinfo {author}
  {\bibfnamefont {B.~A.}\ \bibnamefont {Bernevig}}, \ and\ \bibinfo {author}
  {\bibfnamefont {X.}~\bibnamefont {Dai}},\ }\href {\doibase
  10.1103/PhysRevX.5.011029} {\bibfield  {journal} {\bibinfo  {journal} {Phys.
  Rev. X}\ }\textbf {\bibinfo {volume} {5}},\ \bibinfo {pages} {011029}
  (\bibinfo {year} {2015})}\BibitemShut {NoStop}%
\bibitem [{\citenamefont {Yu}\ \emph {et~al.}(2015)\citenamefont {Yu},
  \citenamefont {Weng}, \citenamefont {Fang}, \citenamefont {Dai},\ and\
  \citenamefont {Hu}}]{Yu2015}%
  \BibitemOpen
  \bibfield  {author} {\bibinfo {author} {\bibfnamefont {R.}~\bibnamefont
  {Yu}}, \bibinfo {author} {\bibfnamefont {H.}~\bibnamefont {Weng}}, \bibinfo
  {author} {\bibfnamefont {Z.}~\bibnamefont {Fang}}, \bibinfo {author}
  {\bibfnamefont {X.}~\bibnamefont {Dai}}, \ and\ \bibinfo {author}
  {\bibfnamefont {X.}~\bibnamefont {Hu}},\ }\href {\doibase
  10.1103/PhysRevLett.115.036807} {\bibfield  {journal} {\bibinfo  {journal}
  {Phys. Rev. Lett.}\ }\textbf {\bibinfo {volume} {115}},\ \bibinfo {pages}
  {036807} (\bibinfo {year} {2015})}\BibitemShut {NoStop}%
\bibitem [{\citenamefont {Xie}\ \emph {et~al.}(2015)\citenamefont {Xie},
  \citenamefont {Schoop}, \citenamefont {Seibel}, \citenamefont {Gibson},
  \citenamefont {Xie},\ and\ \citenamefont {Cava}}]{Xie2015}%
  \BibitemOpen
  \bibfield  {author} {\bibinfo {author} {\bibfnamefont {L.~S.}\ \bibnamefont
  {Xie}}, \bibinfo {author} {\bibfnamefont {L.~M.}\ \bibnamefont {Schoop}},
  \bibinfo {author} {\bibfnamefont {E.~M.}\ \bibnamefont {Seibel}}, \bibinfo
  {author} {\bibfnamefont {Q.~D.}\ \bibnamefont {Gibson}}, \bibinfo {author}
  {\bibfnamefont {W.}~\bibnamefont {Xie}}, \ and\ \bibinfo {author}
  {\bibfnamefont {R.~J.}\ \bibnamefont {Cava}},\ }\href {\doibase
  10.1063/1.4926545} {\bibfield  {journal} {\bibinfo  {journal} {APL
  Materials}\ }\textbf {\bibinfo {volume} {3}},\ \bibinfo {pages} {083602}
  (\bibinfo {year} {2015})}\BibitemShut {NoStop}%
\bibitem [{\citenamefont {Bian}\ \emph {et~al.}(2016)\citenamefont {Bian},
  \citenamefont {Chang}, \citenamefont {Zheng}, \citenamefont {Velury},
  \citenamefont {Xu}, \citenamefont {Neupert}, \citenamefont {Chiu},
  \citenamefont {Huang}, \citenamefont {Sanchez}, \citenamefont {Belopolski},
  \citenamefont {Alidoust}, \citenamefont {Chen}, \citenamefont {Chang},
  \citenamefont {Bansil}, \citenamefont {Jeng}, \citenamefont {Lin},\ and\
  \citenamefont {Hasan}}]{Bian2016}%
  \BibitemOpen
  \bibfield  {author} {\bibinfo {author} {\bibfnamefont {G.}~\bibnamefont
  {Bian}}, \bibinfo {author} {\bibfnamefont {T.-R.}\ \bibnamefont {Chang}},
  \bibinfo {author} {\bibfnamefont {H.}~\bibnamefont {Zheng}}, \bibinfo
  {author} {\bibfnamefont {S.}~\bibnamefont {Velury}}, \bibinfo {author}
  {\bibfnamefont {S.-Y.}\ \bibnamefont {Xu}}, \bibinfo {author} {\bibfnamefont
  {T.}~\bibnamefont {Neupert}}, \bibinfo {author} {\bibfnamefont {C.-K.}\
  \bibnamefont {Chiu}}, \bibinfo {author} {\bibfnamefont {S.-M.}\ \bibnamefont
  {Huang}}, \bibinfo {author} {\bibfnamefont {D.~S.}\ \bibnamefont {Sanchez}},
  \bibinfo {author} {\bibfnamefont {I.}~\bibnamefont {Belopolski}}, \bibinfo
  {author} {\bibfnamefont {N.}~\bibnamefont {Alidoust}}, \bibinfo {author}
  {\bibfnamefont {P.-J.}\ \bibnamefont {Chen}}, \bibinfo {author}
  {\bibfnamefont {G.}~\bibnamefont {Chang}}, \bibinfo {author} {\bibfnamefont
  {A.}~\bibnamefont {Bansil}}, \bibinfo {author} {\bibfnamefont {H.-T.}\
  \bibnamefont {Jeng}}, \bibinfo {author} {\bibfnamefont {H.}~\bibnamefont
  {Lin}}, \ and\ \bibinfo {author} {\bibfnamefont {M.~Z.}\ \bibnamefont
  {Hasan}},\ }\href {\doibase 10.1103/PhysRevB.93.121113} {\bibfield  {journal}
  {\bibinfo  {journal} {Phys. Rev. B}\ }\textbf {\bibinfo {volume} {93}},\
  \bibinfo {pages} {121113} (\bibinfo {year} {2016})}\BibitemShut {NoStop}%
\bibitem [{\citenamefont {Emmanouilidou}\ \emph {et~al.}(2017)\citenamefont
  {Emmanouilidou}, \citenamefont {Shen}, \citenamefont {Deng}, \citenamefont
  {Chang}, \citenamefont {Shi}, \citenamefont {Kotliar}, \citenamefont {Xu},\
  and\ \citenamefont {Ni}}]{nini-2017}%
  \BibitemOpen
  \bibfield  {author} {\bibinfo {author} {\bibfnamefont {E.}~\bibnamefont
  {Emmanouilidou}}, \bibinfo {author} {\bibfnamefont {B.}~\bibnamefont {Shen}},
  \bibinfo {author} {\bibfnamefont {X.}~\bibnamefont {Deng}}, \bibinfo {author}
  {\bibfnamefont {T.-R.}\ \bibnamefont {Chang}}, \bibinfo {author}
  {\bibfnamefont {A.}~\bibnamefont {Shi}}, \bibinfo {author} {\bibfnamefont
  {G.}~\bibnamefont {Kotliar}}, \bibinfo {author} {\bibfnamefont {S.-Y.}\
  \bibnamefont {Xu}}, \ and\ \bibinfo {author} {\bibfnamefont {N.}~\bibnamefont
  {Ni}},\ }\href {\doibase 10.1103/PhysRevB.95.245113} {\bibfield  {journal}
  {\bibinfo  {journal} {Phys. Rev. B}\ }\textbf {\bibinfo {volume} {95}},\
  \bibinfo {pages} {245113} (\bibinfo {year} {2017})}\BibitemShut {NoStop}%
\bibitem [{\citenamefont {Wang}\ \emph {et~al.}(2017)\citenamefont {Wang},
  \citenamefont {Ma}, \citenamefont {Emmanouilidou}, \citenamefont {Shen},
  \citenamefont {Hsu}, \citenamefont {Zhou}, \citenamefont {Zuo}, \citenamefont
  {Song}, \citenamefont {Xu}, \citenamefont {Wang}, \citenamefont {Huang},
  \citenamefont {Ni},\ and\ \citenamefont {Liu}}]{nini2-2017}%
  \BibitemOpen
  \bibfield  {author} {\bibinfo {author} {\bibfnamefont {X.-B.}\ \bibnamefont
  {Wang}}, \bibinfo {author} {\bibfnamefont {X.-M.}\ \bibnamefont {Ma}},
  \bibinfo {author} {\bibfnamefont {E.}~\bibnamefont {Emmanouilidou}}, \bibinfo
  {author} {\bibfnamefont {B.}~\bibnamefont {Shen}}, \bibinfo {author}
  {\bibfnamefont {C.-H.}\ \bibnamefont {Hsu}}, \bibinfo {author} {\bibfnamefont
  {C.-S.}\ \bibnamefont {Zhou}}, \bibinfo {author} {\bibfnamefont
  {Y.}~\bibnamefont {Zuo}}, \bibinfo {author} {\bibfnamefont {R.-R.}\
  \bibnamefont {Song}}, \bibinfo {author} {\bibfnamefont {S.-Y.}\ \bibnamefont
  {Xu}}, \bibinfo {author} {\bibfnamefont {G.}~\bibnamefont {Wang}}, \bibinfo
  {author} {\bibfnamefont {L.}~\bibnamefont {Huang}}, \bibinfo {author}
  {\bibfnamefont {N.}~\bibnamefont {Ni}}, \ and\ \bibinfo {author}
  {\bibfnamefont {C.}~\bibnamefont {Liu}},\ }\href {\doibase
  10.1103/PhysRevB.96.161112} {\bibfield  {journal} {\bibinfo  {journal} {Phys.
  Rev. B}\ }\textbf {\bibinfo {volume} {96}},\ \bibinfo {pages} {161112}
  (\bibinfo {year} {2017})}\BibitemShut {NoStop}%
\bibitem [{\citenamefont {{Zhou}}\ \emph {et~al.}(2017)\citenamefont {{Zhou}},
  \citenamefont {{Liu}}, \citenamefont {{Wu}}, \citenamefont {{Nummy}},
  \citenamefont {{Li}}, \citenamefont {{Griffith}}, \citenamefont {{Parham}},
  \citenamefont {{Waugh}}, \citenamefont {{Emmanouilidou}}, \citenamefont
  {{Shen}}, \citenamefont {{Yazyev}}, \citenamefont {{Ni}},\ and\ \citenamefont
  {{Dessau}}}]{dessau-2017}%
  \BibitemOpen
  \bibfield  {author} {\bibinfo {author} {\bibfnamefont {X.}~\bibnamefont
  {{Zhou}}}, \bibinfo {author} {\bibfnamefont {Q.}~\bibnamefont {{Liu}}},
  \bibinfo {author} {\bibfnamefont {Q.}~\bibnamefont {{Wu}}}, \bibinfo {author}
  {\bibfnamefont {T.}~\bibnamefont {{Nummy}}}, \bibinfo {author} {\bibfnamefont
  {H.}~\bibnamefont {{Li}}}, \bibinfo {author} {\bibfnamefont {J.}~\bibnamefont
  {{Griffith}}}, \bibinfo {author} {\bibfnamefont {S.}~\bibnamefont
  {{Parham}}}, \bibinfo {author} {\bibfnamefont {J.}~\bibnamefont {{Waugh}}},
  \bibinfo {author} {\bibfnamefont {E.}~\bibnamefont {{Emmanouilidou}}},
  \bibinfo {author} {\bibfnamefont {B.}~\bibnamefont {{Shen}}}, \bibinfo
  {author} {\bibfnamefont {O.~V.}\ \bibnamefont {{Yazyev}}}, \bibinfo {author}
  {\bibfnamefont {N.}~\bibnamefont {{Ni}}}, \ and\ \bibinfo {author}
  {\bibfnamefont {D.}~\bibnamefont {{Dessau}}},\ }\href@noop {} {\bibfield
  {journal} {\bibinfo  {journal} {ArXiv e-prints}\ } (\bibinfo {year}
  {2017})},\ \Eprint {http://arxiv.org/abs/1709.10245} {arXiv:1709.10245
  [cond-mat.mtrl-sci]} \BibitemShut {NoStop}%
\bibitem [{\citenamefont {Yamakage}\ \emph {et~al.}(2016)\citenamefont
  {Yamakage}, \citenamefont {Yamakawa}, \citenamefont {Tanaka},\ and\
  \citenamefont {Okamoto}}]{Yamakage2016}%
  \BibitemOpen
  \bibfield  {author} {\bibinfo {author} {\bibfnamefont {A.}~\bibnamefont
  {Yamakage}}, \bibinfo {author} {\bibfnamefont {Y.}~\bibnamefont {Yamakawa}},
  \bibinfo {author} {\bibfnamefont {Y.}~\bibnamefont {Tanaka}}, \ and\ \bibinfo
  {author} {\bibfnamefont {Y.}~\bibnamefont {Okamoto}},\ }\href {\doibase
  10.7566/JPSJ.85.013708} {\bibfield  {journal} {\bibinfo  {journal} {Journal
  of the Physical Society of Japan}\ }\textbf {\bibinfo {volume} {85}},\
  \bibinfo {pages} {013708} (\bibinfo {year} {2016})}\BibitemShut {NoStop}%
\bibitem [{\citenamefont {Okamoto}\ \emph {et~al.}(2016)\citenamefont
  {Okamoto}, \citenamefont {Inohara}, \citenamefont {Yamakage}, \citenamefont
  {Yamakawa},\ and\ \citenamefont {Takenaka}}]{Okamoto2016}%
  \BibitemOpen
  \bibfield  {author} {\bibinfo {author} {\bibfnamefont {Y.}~\bibnamefont
  {Okamoto}}, \bibinfo {author} {\bibfnamefont {T.}~\bibnamefont {Inohara}},
  \bibinfo {author} {\bibfnamefont {A.}~\bibnamefont {Yamakage}}, \bibinfo
  {author} {\bibfnamefont {Y.}~\bibnamefont {Yamakawa}}, \ and\ \bibinfo
  {author} {\bibfnamefont {K.}~\bibnamefont {Takenaka}},\ }\href {\doibase
  10.7566/JPSJ.85.123701} {\bibfield  {journal} {\bibinfo  {journal} {Journal
  of the Physical Society of Japan}\ }\textbf {\bibinfo {volume} {85}},\
  \bibinfo {pages} {123701} (\bibinfo {year} {2016})}\BibitemShut {NoStop}%
\bibitem [{\citenamefont {Takane}\ \emph {et~al.}(2017)\citenamefont {Takane},
  \citenamefont {Nakayama}, \citenamefont {Souma}, \citenamefont {Wada},
  \citenamefont {Okamoto}, \citenamefont {Takenaka}, \citenamefont {Yamakawa},
  \citenamefont {Yamakage}, \citenamefont {Mitsuhashi}, \citenamefont {Horiba},
  \citenamefont {Kumigashira}, \citenamefont {Takahashi},\ and\ \citenamefont
  {Sato}}]{Takane}%
  \BibitemOpen
  \bibfield  {author} {\bibinfo {author} {\bibfnamefont {D.}~\bibnamefont
  {Takane}}, \bibinfo {author} {\bibfnamefont {K.}~\bibnamefont {Nakayama}},
  \bibinfo {author} {\bibfnamefont {S.}~\bibnamefont {Souma}}, \bibinfo
  {author} {\bibfnamefont {T.}~\bibnamefont {Wada}}, \bibinfo {author}
  {\bibfnamefont {Y.}~\bibnamefont {Okamoto}}, \bibinfo {author} {\bibfnamefont
  {K.}~\bibnamefont {Takenaka}}, \bibinfo {author} {\bibfnamefont
  {Y.}~\bibnamefont {Yamakawa}}, \bibinfo {author} {\bibfnamefont
  {A.}~\bibnamefont {Yamakage}}, \bibinfo {author} {\bibfnamefont
  {T.}~\bibnamefont {Mitsuhashi}}, \bibinfo {author} {\bibfnamefont
  {K.}~\bibnamefont {Horiba}}, \bibinfo {author} {\bibfnamefont
  {H.}~\bibnamefont {Kumigashira}}, \bibinfo {author} {\bibfnamefont
  {T.}~\bibnamefont {Takahashi}}, \ and\ \bibinfo {author} {\bibfnamefont
  {T.}~\bibnamefont {Sato}},\ }\href {http://arxiv.org/abs/1708.06874}
  {\bibfield  {journal} {\bibinfo  {journal} {arXiv:1708.06874}\ } (\bibinfo
  {year} {2017})}\BibitemShut {NoStop}%
\bibitem [{\citenamefont {Nayak}\ \emph {et~al.}(2017)\citenamefont {Nayak},
  \citenamefont {Kumar}, \citenamefont {Wu}, \citenamefont {Shekhar},
  \citenamefont {Fink}, \citenamefont {Rienks}, \citenamefont {Fecher},
  \citenamefont {Sun},\ and\ \citenamefont {Felser}}]{Nayak}%
  \BibitemOpen
  \bibfield  {author} {\bibinfo {author} {\bibfnamefont {J.}~\bibnamefont
  {Nayak}}, \bibinfo {author} {\bibfnamefont {N.}~\bibnamefont {Kumar}},
  \bibinfo {author} {\bibfnamefont {S.-C.}\ \bibnamefont {Wu}}, \bibinfo
  {author} {\bibfnamefont {C.}~\bibnamefont {Shekhar}}, \bibinfo {author}
  {\bibfnamefont {J.}~\bibnamefont {Fink}}, \bibinfo {author} {\bibfnamefont
  {E.~E.}\ \bibnamefont {Rienks}}, \bibinfo {author} {\bibfnamefont {G.~H.}\
  \bibnamefont {Fecher}}, \bibinfo {author} {\bibfnamefont {Y.}~\bibnamefont
  {Sun}}, \ and\ \bibinfo {author} {\bibfnamefont {C.}~\bibnamefont {Felser}},\
  }\href {http://arxiv.org/abs/1708.07814} {\bibfield  {journal} {\bibinfo
  {journal} {arXiv:1708.07814}\ } (\bibinfo {year} {2017})}\BibitemShut
  {NoStop}%
\bibitem [{\citenamefont {Heikkil{\"a}}\ and\ \citenamefont
  {Volovik}(2011)}]{Volovik2011}%
  \BibitemOpen
  \bibfield  {author} {\bibinfo {author} {\bibfnamefont {T.~T.}\ \bibnamefont
  {Heikkil{\"a}}}\ and\ \bibinfo {author} {\bibfnamefont {G.~E.}\ \bibnamefont
  {Volovik}},\ }\href {\doibase 10.1134/S002136401102007X} {\bibfield
  {journal} {\bibinfo  {journal} {JETP Letters}\ }\textbf {\bibinfo {volume}
  {93}},\ \bibinfo {pages} {59} (\bibinfo {year} {2011})}\BibitemShut {NoStop}%
\bibitem [{\citenamefont {Burkov}\ \emph {et~al.}(2011)\citenamefont {Burkov},
  \citenamefont {Hook},\ and\ \citenamefont {Balents}}]{Burkov2011}%
  \BibitemOpen
  \bibfield  {author} {\bibinfo {author} {\bibfnamefont {A.~A.}\ \bibnamefont
  {Burkov}}, \bibinfo {author} {\bibfnamefont {M.~D.}\ \bibnamefont {Hook}}, \
  and\ \bibinfo {author} {\bibfnamefont {L.}~\bibnamefont {Balents}},\ }\href
  {\doibase 10.1103/PhysRevB.84.235126} {\bibfield  {journal} {\bibinfo
  {journal} {Phys. Rev. B}\ }\textbf {\bibinfo {volume} {84}},\ \bibinfo
  {pages} {235126} (\bibinfo {year} {2011})}\BibitemShut {NoStop}%
\bibitem [{\citenamefont {Chan}\ \emph {et~al.}(2016)\citenamefont {Chan},
  \citenamefont {Chiu}, \citenamefont {Chou},\ and\ \citenamefont
  {Schnyder}}]{Chan2016}%
  \BibitemOpen
  \bibfield  {author} {\bibinfo {author} {\bibfnamefont {Y.-H.}\ \bibnamefont
  {Chan}}, \bibinfo {author} {\bibfnamefont {C.-K.}\ \bibnamefont {Chiu}},
  \bibinfo {author} {\bibfnamefont {M.~Y.}\ \bibnamefont {Chou}}, \ and\
  \bibinfo {author} {\bibfnamefont {A.~P.}\ \bibnamefont {Schnyder}},\ }\href
  {\doibase 10.1103/PhysRevB.93.205132} {\bibfield  {journal} {\bibinfo
  {journal} {Phys. Rev. B}\ }\textbf {\bibinfo {volume} {93}},\ \bibinfo
  {pages} {205132} (\bibinfo {year} {2016})}\BibitemShut {NoStop}%
\bibitem [{\citenamefont {Ramamurthy}\ and\ \citenamefont
  {Hughes}(2017)}]{Ramamurthy2017}%
  \BibitemOpen
  \bibfield  {author} {\bibinfo {author} {\bibfnamefont {S.~T.}\ \bibnamefont
  {Ramamurthy}}\ and\ \bibinfo {author} {\bibfnamefont {T.~L.}\ \bibnamefont
  {Hughes}},\ }\href {\doibase 10.1103/PhysRevB.95.075138} {\bibfield
  {journal} {\bibinfo  {journal} {Phys. Rev. B}\ }\textbf {\bibinfo {volume}
  {95}},\ \bibinfo {pages} {075138} (\bibinfo {year} {2017})}\BibitemShut
  {NoStop}%
\bibitem [{\citenamefont {Rhim}\ and\ \citenamefont {Kim}(2015)}]{Rhim2015}%
  \BibitemOpen
  \bibfield  {author} {\bibinfo {author} {\bibfnamefont {J.-W.}\ \bibnamefont
  {Rhim}}\ and\ \bibinfo {author} {\bibfnamefont {Y.~B.}\ \bibnamefont {Kim}},\
  }\href {\doibase 10.1103/PhysRevB.92.045126} {\bibfield  {journal} {\bibinfo
  {journal} {Phys. Rev. B}\ }\textbf {\bibinfo {volume} {92}},\ \bibinfo
  {pages} {045126} (\bibinfo {year} {2015})}\BibitemShut {NoStop}%
\bibitem [{\citenamefont {Liu}\ and\ \citenamefont {Balents}(2017)}]{Liu2017}%
  \BibitemOpen
  \bibfield  {author} {\bibinfo {author} {\bibfnamefont {J.}~\bibnamefont
  {Liu}}\ and\ \bibinfo {author} {\bibfnamefont {L.}~\bibnamefont {Balents}},\
  }\href {\doibase 10.1103/PhysRevB.95.075426} {\bibfield  {journal} {\bibinfo
  {journal} {Phys. Rev. B}\ }\textbf {\bibinfo {volume} {95}},\ \bibinfo
  {pages} {075426} (\bibinfo {year} {2017})}\BibitemShut {NoStop}%
\bibitem [{\citenamefont {Nandkishore}(2016)}]{Nandkishore2016}%
  \BibitemOpen
  \bibfield  {author} {\bibinfo {author} {\bibfnamefont {R.}~\bibnamefont
  {Nandkishore}},\ }\href {\doibase 10.1103/PhysRevB.93.020506} {\bibfield
  {journal} {\bibinfo  {journal} {Phys. Rev. B}\ }\textbf {\bibinfo {volume}
  {93}},\ \bibinfo {pages} {020506} (\bibinfo {year} {2016})}\BibitemShut
  {NoStop}%
\bibitem [{\citenamefont {Wang}\ and\ \citenamefont {Ye}(2016)}]{Wang-Ye2016}%
  \BibitemOpen
  \bibfield  {author} {\bibinfo {author} {\bibfnamefont {Y.}~\bibnamefont
  {Wang}}\ and\ \bibinfo {author} {\bibfnamefont {P.}~\bibnamefont {Ye}},\
  }\href {\doibase 10.1103/PhysRevB.94.075115} {\bibfield  {journal} {\bibinfo
  {journal} {Phys. Rev. B}\ }\textbf {\bibinfo {volume} {94}},\ \bibinfo
  {pages} {075115} (\bibinfo {year} {2016})}\BibitemShut {NoStop}%
\bibitem [{\citenamefont {Roy}(2017)}]{Roy2017}%
  \BibitemOpen
  \bibfield  {author} {\bibinfo {author} {\bibfnamefont {B.}~\bibnamefont
  {Roy}},\ }\href {\doibase 10.1103/PhysRevB.96.041113} {\bibfield  {journal}
  {\bibinfo  {journal} {Phys. Rev. B}\ }\textbf {\bibinfo {volume} {96}},\
  \bibinfo {pages} {041113} (\bibinfo {year} {2017})}\BibitemShut {NoStop}%
\bibitem [{\citenamefont {Wang}\ and\ \citenamefont
  {Nandkishore}(2017{\natexlab{a}})}]{wang-nandkishore2-2017}%
  \BibitemOpen
  \bibfield  {author} {\bibinfo {author} {\bibfnamefont {Y.}~\bibnamefont
  {Wang}}\ and\ \bibinfo {author} {\bibfnamefont {R.~M.}\ \bibnamefont
  {Nandkishore}},\ }\href {\doibase 10.1103/PhysRevB.96.115130} {\bibfield
  {journal} {\bibinfo  {journal} {Phys. Rev. B}\ }\textbf {\bibinfo {volume}
  {96}},\ \bibinfo {pages} {115130} (\bibinfo {year}
  {2017}{\natexlab{a}})}\BibitemShut {NoStop}%
\bibitem [{\citenamefont {Guan}\ \emph {et~al.}(2016)\citenamefont {Guan},
  \citenamefont {Chen}, \citenamefont {Chu}, \citenamefont {Sankar},
  \citenamefont {Chou}, \citenamefont {Jeng}, \citenamefont {Chang},\ and\
  \citenamefont {Chuang}}]{Guan-2016}%
  \BibitemOpen
  \bibfield  {author} {\bibinfo {author} {\bibfnamefont {S.-Y.}\ \bibnamefont
  {Guan}}, \bibinfo {author} {\bibfnamefont {P.-J.}\ \bibnamefont {Chen}},
  \bibinfo {author} {\bibfnamefont {M.-W.}\ \bibnamefont {Chu}}, \bibinfo
  {author} {\bibfnamefont {R.}~\bibnamefont {Sankar}}, \bibinfo {author}
  {\bibfnamefont {F.}~\bibnamefont {Chou}}, \bibinfo {author} {\bibfnamefont
  {H.-T.}\ \bibnamefont {Jeng}}, \bibinfo {author} {\bibfnamefont {C.-S.}\
  \bibnamefont {Chang}}, \ and\ \bibinfo {author} {\bibfnamefont {T.-M.}\
  \bibnamefont {Chuang}},\ }\href {\doibase 10.1126/sciadv.1600894} {\bibfield
  {journal} {\bibinfo  {journal} {Science Advances}\ }\textbf {\bibinfo
  {volume} {2}} (\bibinfo {year} {2016}),\ 10.1126/sciadv.1600894},\ \Eprint
  {http://arxiv.org/abs/http://advances.sciencemag.org/content/2/11/e1600894.full.pdf}
  {http://advances.sciencemag.org/content/2/11/e1600894.full.pdf} \BibitemShut
  {NoStop}%
\bibitem [{\citenamefont {Sur}\ and\ \citenamefont
  {Nandkishore}(2016)}]{Sur-Nandkishore2016}%
  \BibitemOpen
  \bibfield  {author} {\bibinfo {author} {\bibfnamefont {S.}~\bibnamefont
  {Sur}}\ and\ \bibinfo {author} {\bibfnamefont {R.}~\bibnamefont
  {Nandkishore}},\ }\href {http://stacks.iop.org/1367-2630/18/i=11/a=115006}
  {\bibfield  {journal} {\bibinfo  {journal} {New Journal of Physics}\ }\textbf
  {\bibinfo {volume} {18}},\ \bibinfo {pages} {115006} (\bibinfo {year}
  {2016})}\BibitemShut {NoStop}%
\bibitem [{\citenamefont {Wang}\ and\ \citenamefont
  {Nandkishore}(2017{\natexlab{b}})}]{Wang-Nandkishore2017}%
  \BibitemOpen
  \bibfield  {author} {\bibinfo {author} {\bibfnamefont {Y.}~\bibnamefont
  {Wang}}\ and\ \bibinfo {author} {\bibfnamefont {R.~M.}\ \bibnamefont
  {Nandkishore}},\ }\href {\doibase 10.1103/PhysRevB.95.060506} {\bibfield
  {journal} {\bibinfo  {journal} {Phys. Rev. B}\ }\textbf {\bibinfo {volume}
  {95}},\ \bibinfo {pages} {060506} (\bibinfo {year}
  {2017}{\natexlab{b}})}\BibitemShut {NoStop}%
\bibitem [{\citenamefont {Altland}\ and\ \citenamefont
  {Zirnbauer}(1997)}]{AZ_1997}%
  \BibitemOpen
  \bibfield  {author} {\bibinfo {author} {\bibfnamefont {A.}~\bibnamefont
  {Altland}}\ and\ \bibinfo {author} {\bibfnamefont {M.~R.}\ \bibnamefont
  {Zirnbauer}},\ }\href {\doibase 10.1103/PhysRevB.55.1142} {\bibfield
  {journal} {\bibinfo  {journal} {Phys. Rev. B}\ }\textbf {\bibinfo {volume}
  {55}},\ \bibinfo {pages} {1142} (\bibinfo {year} {1997})}\BibitemShut
  {NoStop}%
\bibitem [{\citenamefont {Schnyder}\ \emph {et~al.}(2008)\citenamefont
  {Schnyder}, \citenamefont {Ryu}, \citenamefont {Furusaki},\ and\
  \citenamefont {Ludwig}}]{Schnyder_class}%
  \BibitemOpen
  \bibfield  {author} {\bibinfo {author} {\bibfnamefont {A.~P.}\ \bibnamefont
  {Schnyder}}, \bibinfo {author} {\bibfnamefont {S.}~\bibnamefont {Ryu}},
  \bibinfo {author} {\bibfnamefont {A.}~\bibnamefont {Furusaki}}, \ and\
  \bibinfo {author} {\bibfnamefont {A.~W.~W.}\ \bibnamefont {Ludwig}},\ }\href
  {\doibase 10.1103/PhysRevB.78.195125} {\bibfield  {journal} {\bibinfo
  {journal} {Phys. Rev. B}\ }\textbf {\bibinfo {volume} {78}},\ \bibinfo
  {pages} {195125} (\bibinfo {year} {2008})}\BibitemShut {NoStop}%
\bibitem [{\citenamefont {Kitaev}(2009)}]{Kitaev_class}%
  \BibitemOpen
  \bibfield  {author} {\bibinfo {author} {\bibfnamefont {A.}~\bibnamefont
  {Kitaev}},\ }\href {\doibase 10.1063/1.3149495} {\bibfield  {journal}
  {\bibinfo  {journal} {AIP Conference Proceedings}\ }\textbf {\bibinfo
  {volume} {1134}},\ \bibinfo {pages} {22} (\bibinfo {year}
  {2009})}\BibitemShut {NoStop}%
\bibitem [{\citenamefont {Ryu}\ \emph {et~al.}(2010)\citenamefont {Ryu},
  \citenamefont {Schnyder}, \citenamefont {Furusaki},\ and\ \citenamefont
  {Ludwig}}]{Ryu_class}%
  \BibitemOpen
  \bibfield  {author} {\bibinfo {author} {\bibfnamefont {S.}~\bibnamefont
  {Ryu}}, \bibinfo {author} {\bibfnamefont {A.~P.}\ \bibnamefont {Schnyder}},
  \bibinfo {author} {\bibfnamefont {A.}~\bibnamefont {Furusaki}}, \ and\
  \bibinfo {author} {\bibfnamefont {A.~W.~W.}\ \bibnamefont {Ludwig}},\ }\href
  {http://stacks.iop.org/1367-2630/12/i=6/a=065010} {\bibfield  {journal}
  {\bibinfo  {journal} {New Journal of Physics}\ }\textbf {\bibinfo {volume}
  {12}},\ \bibinfo {pages} {065010} (\bibinfo {year} {2010})}\BibitemShut
  {NoStop}%
\bibitem [{\citenamefont {Chiu}\ \emph {et~al.}(2013)\citenamefont {Chiu},
  \citenamefont {Yao},\ and\ \citenamefont {Ryu}}]{Chiu2013}%
  \BibitemOpen
  \bibfield  {author} {\bibinfo {author} {\bibfnamefont {C.-K.}\ \bibnamefont
  {Chiu}}, \bibinfo {author} {\bibfnamefont {H.}~\bibnamefont {Yao}}, \ and\
  \bibinfo {author} {\bibfnamefont {S.}~\bibnamefont {Ryu}},\ }\href {\doibase
  10.1103/PhysRevB.88.075142} {\bibfield  {journal} {\bibinfo  {journal} {Phys.
  Rev. B}\ }\textbf {\bibinfo {volume} {88}},\ \bibinfo {pages} {075142}
  (\bibinfo {year} {2013})}\BibitemShut {NoStop}%
\bibitem [{\citenamefont {Morimoto}\ and\ \citenamefont
  {Furusaki}(2013)}]{Morimoto2013}%
  \BibitemOpen
  \bibfield  {author} {\bibinfo {author} {\bibfnamefont {T.}~\bibnamefont
  {Morimoto}}\ and\ \bibinfo {author} {\bibfnamefont {A.}~\bibnamefont
  {Furusaki}},\ }\href {\doibase 10.1103/PhysRevB.88.125129} {\bibfield
  {journal} {\bibinfo  {journal} {Phys. Rev. B}\ }\textbf {\bibinfo {volume}
  {88}},\ \bibinfo {pages} {125129} (\bibinfo {year} {2013})}\BibitemShut
  {NoStop}%
\bibitem [{\citenamefont {Shiozaki}\ and\ \citenamefont
  {Sato}(2014)}]{Shiozaki2014}%
  \BibitemOpen
  \bibfield  {author} {\bibinfo {author} {\bibfnamefont {K.}~\bibnamefont
  {Shiozaki}}\ and\ \bibinfo {author} {\bibfnamefont {M.}~\bibnamefont
  {Sato}},\ }\href {\doibase 10.1103/PhysRevB.90.165114} {\bibfield  {journal}
  {\bibinfo  {journal} {Phys. Rev. B}\ }\textbf {\bibinfo {volume} {90}},\
  \bibinfo {pages} {165114} (\bibinfo {year} {2014})}\BibitemShut {NoStop}%
\bibitem [{\citenamefont {Chiu}\ \emph {et~al.}(2016)\citenamefont {Chiu},
  \citenamefont {Teo}, \citenamefont {Schnyder},\ and\ \citenamefont
  {Ryu}}]{Chiu_RMP}%
  \BibitemOpen
  \bibfield  {author} {\bibinfo {author} {\bibfnamefont {C.-K.}\ \bibnamefont
  {Chiu}}, \bibinfo {author} {\bibfnamefont {J.~C.~Y.}\ \bibnamefont {Teo}},
  \bibinfo {author} {\bibfnamefont {A.~P.}\ \bibnamefont {Schnyder}}, \ and\
  \bibinfo {author} {\bibfnamefont {S.}~\bibnamefont {Ryu}},\ }\href {\doibase
  10.1103/RevModPhys.88.035005} {\bibfield  {journal} {\bibinfo  {journal}
  {Rev. Mod. Phys.}\ }\textbf {\bibinfo {volume} {88}},\ \bibinfo {pages}
  {035005} (\bibinfo {year} {2016})}\BibitemShut {NoStop}%
\bibitem [{\citenamefont {Essin}\ \emph {et~al.}(2009)\citenamefont {Essin},
  \citenamefont {Moore},\ and\ \citenamefont {Vanderbilt}}]{Essin-Moore}%
  \BibitemOpen
  \bibfield  {author} {\bibinfo {author} {\bibfnamefont {A.~M.}\ \bibnamefont
  {Essin}}, \bibinfo {author} {\bibfnamefont {J.~E.}\ \bibnamefont {Moore}}, \
  and\ \bibinfo {author} {\bibfnamefont {D.}~\bibnamefont {Vanderbilt}},\
  }\href {\doibase 10.1103/PhysRevLett.102.146805} {\bibfield  {journal}
  {\bibinfo  {journal} {Phys. Rev. Lett.}\ }\textbf {\bibinfo {volume} {102}},\
  \bibinfo {pages} {146805} (\bibinfo {year} {2009})}\BibitemShut {NoStop}%
\bibitem [{\citenamefont {Benalcazar}\ \emph
  {et~al.}(2017{\natexlab{a}})\citenamefont {Benalcazar}, \citenamefont
  {Bernevig},\ and\ \citenamefont {Hughes}}]{Hughes1}%
  \BibitemOpen
  \bibfield  {author} {\bibinfo {author} {\bibfnamefont {W.~A.}\ \bibnamefont
  {Benalcazar}}, \bibinfo {author} {\bibfnamefont {B.~A.}\ \bibnamefont
  {Bernevig}}, \ and\ \bibinfo {author} {\bibfnamefont {T.~L.}\ \bibnamefont
  {Hughes}},\ }\href {\doibase 10.1126/science.aah6442} {\bibfield  {journal}
  {\bibinfo  {journal} {Science}\ }\textbf {\bibinfo {volume} {357}},\ \bibinfo
  {pages} {61} (\bibinfo {year} {2017}{\natexlab{a}})}\BibitemShut {NoStop}%
\bibitem [{\citenamefont {Benalcazar}\ \emph
  {et~al.}(2017{\natexlab{b}})\citenamefont {Benalcazar}, \citenamefont
  {Bernevig},\ and\ \citenamefont {Hughes}}]{Hughes2}%
  \BibitemOpen
  \bibfield  {author} {\bibinfo {author} {\bibfnamefont {W.~A.}\ \bibnamefont
  {Benalcazar}}, \bibinfo {author} {\bibfnamefont {B.~A.}\ \bibnamefont
  {Bernevig}}, \ and\ \bibinfo {author} {\bibfnamefont {T.~L.}\ \bibnamefont
  {Hughes}},\ }\href {http://arxiv.org/abs/1708.04230} {\bibfield  {journal}
  {\bibinfo  {journal} {arXiv:1708.04230}\ } (\bibinfo {year}
  {2017}{\natexlab{b}})}\BibitemShut {NoStop}%
\bibitem [{\citenamefont {Song}\ \emph {et~al.}(2017)\citenamefont {Song},
  \citenamefont {Fang},\ and\ \citenamefont {Fang}}]{Fang}%
  \BibitemOpen
  \bibfield  {author} {\bibinfo {author} {\bibfnamefont {Z.}~\bibnamefont
  {Song}}, \bibinfo {author} {\bibfnamefont {Z.}~\bibnamefont {Fang}}, \ and\
  \bibinfo {author} {\bibfnamefont {C.}~\bibnamefont {Fang}},\ }\href
  {http://arxiv.org/abs/1708.02952} {\bibfield  {journal} {\bibinfo  {journal}
  {arXiv:1708.02952}\ } (\bibinfo {year} {2017})}\BibitemShut {NoStop}%
\bibitem [{\citenamefont {Schindler}\ \emph {et~al.}(2017)\citenamefont
  {Schindler}, \citenamefont {Cook}, \citenamefont {Vergniory}, \citenamefont
  {Wang}, \citenamefont {Parkin}, \citenamefont {Bernevig},\ and\ \citenamefont
  {Neupert}}]{Neupert}%
  \BibitemOpen
  \bibfield  {author} {\bibinfo {author} {\bibfnamefont {F.}~\bibnamefont
  {Schindler}}, \bibinfo {author} {\bibfnamefont {A.~M.}\ \bibnamefont {Cook}},
  \bibinfo {author} {\bibfnamefont {M.~G.}\ \bibnamefont {Vergniory}}, \bibinfo
  {author} {\bibfnamefont {Z.}~\bibnamefont {Wang}}, \bibinfo {author}
  {\bibfnamefont {S.~S.~P.}\ \bibnamefont {Parkin}}, \bibinfo {author}
  {\bibfnamefont {B.~A.}\ \bibnamefont {Bernevig}}, \ and\ \bibinfo {author}
  {\bibfnamefont {T.}~\bibnamefont {Neupert}},\ }\href
  {http://arxiv.org/abs/1708.03636} {\bibfield  {journal} {\bibinfo  {journal}
  {arXiv:1708.03636}\ } (\bibinfo {year} {2017})}\BibitemShut {NoStop}%
\bibitem [{\citenamefont {Langbehn}\ \emph {et~al.}(2017)\citenamefont
  {Langbehn}, \citenamefont {Peng}, \citenamefont {Trifunovic}, \citenamefont
  {von Oppen},\ and\ \citenamefont {Brouwer}}]{Luka}%
  \BibitemOpen
  \bibfield  {author} {\bibinfo {author} {\bibfnamefont {J.}~\bibnamefont
  {Langbehn}}, \bibinfo {author} {\bibfnamefont {Y.}~\bibnamefont {Peng}},
  \bibinfo {author} {\bibfnamefont {L.}~\bibnamefont {Trifunovic}}, \bibinfo
  {author} {\bibfnamefont {F.}~\bibnamefont {von Oppen}}, \ and\ \bibinfo
  {author} {\bibfnamefont {P.~W.}\ \bibnamefont {Brouwer}},\ }\href
  {http://arxiv.org/abs/1708.03640} {\bibfield  {journal} {\bibinfo  {journal}
  {arXiv:1708.03640}\ } (\bibinfo {year} {2017})}\BibitemShut {NoStop}%
\bibitem [{\citenamefont {{Schindler}}\ \emph {et~al.}(2018)\citenamefont
  {{Schindler}}, \citenamefont {{Wang}}, \citenamefont {{Vergniory}},
  \citenamefont {{Cook}}, \citenamefont {{Murani}}, \citenamefont {{Sengupta}},
  \citenamefont {{Kasumov}}, \citenamefont {{Deblock}}, \citenamefont {{Jeon}},
  \citenamefont {{Drozdov}}, \citenamefont {{Bouchiat}}, \citenamefont
  {{Gu{\'e}ron}}, \citenamefont {{Yazdani}}, \citenamefont {{Bernevig}},\ and\
  \citenamefont {{Neupert}}}]{neupert_new}%
  \BibitemOpen
  \bibfield  {author} {\bibinfo {author} {\bibfnamefont {F.}~\bibnamefont
  {{Schindler}}}, \bibinfo {author} {\bibfnamefont {Z.}~\bibnamefont {{Wang}}},
  \bibinfo {author} {\bibfnamefont {M.~G.}\ \bibnamefont {{Vergniory}}},
  \bibinfo {author} {\bibfnamefont {A.~M.}\ \bibnamefont {{Cook}}}, \bibinfo
  {author} {\bibfnamefont {A.}~\bibnamefont {{Murani}}}, \bibinfo {author}
  {\bibfnamefont {S.}~\bibnamefont {{Sengupta}}}, \bibinfo {author}
  {\bibfnamefont {A.~Y.}\ \bibnamefont {{Kasumov}}}, \bibinfo {author}
  {\bibfnamefont {R.}~\bibnamefont {{Deblock}}}, \bibinfo {author}
  {\bibfnamefont {S.}~\bibnamefont {{Jeon}}}, \bibinfo {author} {\bibfnamefont
  {I.}~\bibnamefont {{Drozdov}}}, \bibinfo {author} {\bibfnamefont
  {H.}~\bibnamefont {{Bouchiat}}}, \bibinfo {author} {\bibfnamefont
  {S.}~\bibnamefont {{Gu{\'e}ron}}}, \bibinfo {author} {\bibfnamefont
  {A.}~\bibnamefont {{Yazdani}}}, \bibinfo {author} {\bibfnamefont {B.~A.}\
  \bibnamefont {{Bernevig}}}, \ and\ \bibinfo {author} {\bibfnamefont
  {T.}~\bibnamefont {{Neupert}}},\ }\href@noop {} {\bibfield  {journal}
  {\bibinfo  {journal} {ArXiv e-prints}\ } (\bibinfo {year} {2018})},\ \Eprint
  {http://arxiv.org/abs/1802.02585} {arXiv:1802.02585 [cond-mat.mtrl-sci]}
  \BibitemShut {NoStop}%
\bibitem [{\citenamefont {Read}\ and\ \citenamefont
  {Green}(2000)}]{ReadGreen2000}%
  \BibitemOpen
  \bibfield  {author} {\bibinfo {author} {\bibfnamefont {N.}~\bibnamefont
  {Read}}\ and\ \bibinfo {author} {\bibfnamefont {D.}~\bibnamefont {Green}},\
  }\href {\doibase 10.1103/PhysRevB.61.10267} {\bibfield  {journal} {\bibinfo
  {journal} {Phys. Rev. B}\ }\textbf {\bibinfo {volume} {61}},\ \bibinfo
  {pages} {10267} (\bibinfo {year} {2000})}\BibitemShut {NoStop}%
\bibitem [{\citenamefont {Thouless}(1974)}]{IPR_Thouless}%
  \BibitemOpen
  \bibfield  {author} {\bibinfo {author} {\bibfnamefont {D.}~\bibnamefont
  {Thouless}},\ }\href {\doibase https://doi.org/10.1016/0370-1573(74)90029-5}
  {\bibfield  {journal} {\bibinfo  {journal} {Physics Reports}\ }\textbf
  {\bibinfo {volume} {13}},\ \bibinfo {pages} {93 } (\bibinfo {year}
  {1974})}\BibitemShut {NoStop}%
\bibitem [{\citenamefont {Teo}\ \emph {et~al.}(2008)\citenamefont {Teo},
  \citenamefont {Fu},\ and\ \citenamefont {Kane}}]{fu-teo-kane}%
  \BibitemOpen
  \bibfield  {author} {\bibinfo {author} {\bibfnamefont {J.~C.~Y.}\
  \bibnamefont {Teo}}, \bibinfo {author} {\bibfnamefont {L.}~\bibnamefont
  {Fu}}, \ and\ \bibinfo {author} {\bibfnamefont {C.~L.}\ \bibnamefont
  {Kane}},\ }\href {\doibase 10.1103/PhysRevB.78.045426} {\bibfield  {journal}
  {\bibinfo  {journal} {Phys. Rev. B}\ }\textbf {\bibinfo {volume} {78}},\
  \bibinfo {pages} {045426} (\bibinfo {year} {2008})}\BibitemShut {NoStop}%
\bibitem [{\citenamefont {Thouless}\ \emph {et~al.}(1982)\citenamefont
  {Thouless}, \citenamefont {Kohmoto}, \citenamefont {Nightingale},\ and\
  \citenamefont {den Nijs}}]{TKNN1982}%
  \BibitemOpen
  \bibfield  {author} {\bibinfo {author} {\bibfnamefont {D.~J.}\ \bibnamefont
  {Thouless}}, \bibinfo {author} {\bibfnamefont {M.}~\bibnamefont {Kohmoto}},
  \bibinfo {author} {\bibfnamefont {M.~P.}\ \bibnamefont {Nightingale}}, \ and\
  \bibinfo {author} {\bibfnamefont {M.}~\bibnamefont {den Nijs}},\ }\href
  {\doibase 10.1103/PhysRevLett.49.405} {\bibfield  {journal} {\bibinfo
  {journal} {Phys. Rev. Lett.}\ }\textbf {\bibinfo {volume} {49}},\ \bibinfo
  {pages} {405} (\bibinfo {year} {1982})}\BibitemShut {NoStop}%
\bibitem [{\citenamefont {Alicea}(2012)}]{Alicea2012}%
  \BibitemOpen
  \bibfield  {author} {\bibinfo {author} {\bibfnamefont {J.}~\bibnamefont
  {Alicea}},\ }\href {http://stacks.iop.org/0034-4885/75/i=7/a=076501}
  {\bibfield  {journal} {\bibinfo  {journal} {Reports on Progress in Physics}\
  }\textbf {\bibinfo {volume} {75}},\ \bibinfo {pages} {076501} (\bibinfo
  {year} {2012})}\BibitemShut {NoStop}%
\bibitem [{\citenamefont {Volovik}(1990)}]{Volovik}%
  \BibitemOpen
  \bibfield  {author} {\bibinfo {author} {\bibfnamefont {G.~E.}\ \bibnamefont
  {Volovik}},\ }\href@noop {} {\bibfield  {journal} {\bibinfo  {journal} {JETP
  Lett.}\ }\textbf {\bibinfo {volume} {51}},\ \bibinfo {pages} {125} (\bibinfo
  {year} {1990})}\BibitemShut {NoStop}%
\bibitem [{\citenamefont {Shapourian}\ \emph {et~al.}(2015)\citenamefont
  {Shapourian}, \citenamefont {Hughes},\ and\ \citenamefont
  {Ryu}}]{Shapourian2015}%
  \BibitemOpen
  \bibfield  {author} {\bibinfo {author} {\bibfnamefont {H.}~\bibnamefont
  {Shapourian}}, \bibinfo {author} {\bibfnamefont {T.~L.}\ \bibnamefont
  {Hughes}}, \ and\ \bibinfo {author} {\bibfnamefont {S.}~\bibnamefont {Ryu}},\
  }\href {\doibase 10.1103/PhysRevB.92.165131} {\bibfield  {journal} {\bibinfo
  {journal} {Phys. Rev. B}\ }\textbf {\bibinfo {volume} {92}},\ \bibinfo
  {pages} {165131} (\bibinfo {year} {2015})}\BibitemShut {NoStop}%
\bibitem [{\citenamefont {Qi}\ \emph {et~al.}(2008)\citenamefont {Qi},
  \citenamefont {Hughes},\ and\ \citenamefont {Zhang}}]{Qi_3DTI}%
  \BibitemOpen
  \bibfield  {author} {\bibinfo {author} {\bibfnamefont {X.-L.}\ \bibnamefont
  {Qi}}, \bibinfo {author} {\bibfnamefont {T.~L.}\ \bibnamefont {Hughes}}, \
  and\ \bibinfo {author} {\bibfnamefont {S.-C.}\ \bibnamefont {Zhang}},\ }\href
  {\doibase 10.1103/PhysRevB.78.195424} {\bibfield  {journal} {\bibinfo
  {journal} {Phys. Rev. B}\ }\textbf {\bibinfo {volume} {78}},\ \bibinfo
  {pages} {195424} (\bibinfo {year} {2008})}\BibitemShut {NoStop}%
\bibitem [{\citenamefont {Ryu}\ \emph {et~al.}(2012)\citenamefont {Ryu},
  \citenamefont {Moore},\ and\ \citenamefont {Ludwig}}]{Ryu2012}%
  \BibitemOpen
  \bibfield  {author} {\bibinfo {author} {\bibfnamefont {S.}~\bibnamefont
  {Ryu}}, \bibinfo {author} {\bibfnamefont {J.~E.}\ \bibnamefont {Moore}}, \
  and\ \bibinfo {author} {\bibfnamefont {A.~W.~W.}\ \bibnamefont {Ludwig}},\
  }\href {\doibase 10.1103/PhysRevB.85.045104} {\bibfield  {journal} {\bibinfo
  {journal} {Phys. Rev. B}\ }\textbf {\bibinfo {volume} {85}},\ \bibinfo
  {pages} {045104} (\bibinfo {year} {2012})}\BibitemShut {NoStop}%
\bibitem [{\citenamefont {Nomura}\ \emph {et~al.}(2012)\citenamefont {Nomura},
  \citenamefont {Ryu}, \citenamefont {Furusaki},\ and\ \citenamefont
  {Nagaosa}}]{Nomura2012}%
  \BibitemOpen
  \bibfield  {author} {\bibinfo {author} {\bibfnamefont {K.}~\bibnamefont
  {Nomura}}, \bibinfo {author} {\bibfnamefont {S.}~\bibnamefont {Ryu}},
  \bibinfo {author} {\bibfnamefont {A.}~\bibnamefont {Furusaki}}, \ and\
  \bibinfo {author} {\bibfnamefont {N.}~\bibnamefont {Nagaosa}},\ }\href
  {\doibase 10.1103/PhysRevLett.108.026802} {\bibfield  {journal} {\bibinfo
  {journal} {Phys. Rev. Lett.}\ }\textbf {\bibinfo {volume} {108}},\ \bibinfo
  {pages} {026802} (\bibinfo {year} {2012})}\BibitemShut {NoStop}%
\bibitem [{\citenamefont {Fu}\ and\ \citenamefont {Kane}(2008)}]{fu-kane-2008}%
  \BibitemOpen
  \bibfield  {author} {\bibinfo {author} {\bibfnamefont {L.}~\bibnamefont
  {Fu}}\ and\ \bibinfo {author} {\bibfnamefont {C.~L.}\ \bibnamefont {Kane}},\
  }\href {\doibase 10.1103/PhysRevLett.100.096407} {\bibfield  {journal}
  {\bibinfo  {journal} {Phys. Rev. Lett.}\ }\textbf {\bibinfo {volume} {100}},\
  \bibinfo {pages} {096407} (\bibinfo {year} {2008})}\BibitemShut {NoStop}%
\bibitem [{\citenamefont {Hosur}\ \emph {et~al.}(2011)\citenamefont {Hosur},
  \citenamefont {Ghaemi}, \citenamefont {Mong},\ and\ \citenamefont
  {Vishwanath}}]{hosur-prl}%
  \BibitemOpen
  \bibfield  {author} {\bibinfo {author} {\bibfnamefont {P.}~\bibnamefont
  {Hosur}}, \bibinfo {author} {\bibfnamefont {P.}~\bibnamefont {Ghaemi}},
  \bibinfo {author} {\bibfnamefont {R.~S.~K.}\ \bibnamefont {Mong}}, \ and\
  \bibinfo {author} {\bibfnamefont {A.}~\bibnamefont {Vishwanath}},\ }\href
  {\doibase 10.1103/PhysRevLett.107.097001} {\bibfield  {journal} {\bibinfo
  {journal} {Phys. Rev. Lett.}\ }\textbf {\bibinfo {volume} {107}},\ \bibinfo
  {pages} {097001} (\bibinfo {year} {2011})}\BibitemShut {NoStop}%
\bibitem [{\citenamefont {Ivanov}(2001)}]{ivanov}%
  \BibitemOpen
  \bibfield  {author} {\bibinfo {author} {\bibfnamefont {D.~A.}\ \bibnamefont
  {Ivanov}},\ }\href {\doibase 10.1103/PhysRevLett.86.268} {\bibfield
  {journal} {\bibinfo  {journal} {Phys. Rev. Lett.}\ }\textbf {\bibinfo
  {volume} {86}},\ \bibinfo {pages} {268} (\bibinfo {year} {2001})}\BibitemShut
  {NoStop}%
\bibitem [{\citenamefont {Nishida}\ \emph {et~al.}(2010)\citenamefont
  {Nishida}, \citenamefont {Santos},\ and\ \citenamefont
  {Chamon}}]{Nishida2010}%
  \BibitemOpen
  \bibfield  {author} {\bibinfo {author} {\bibfnamefont {Y.}~\bibnamefont
  {Nishida}}, \bibinfo {author} {\bibfnamefont {L.}~\bibnamefont {Santos}}, \
  and\ \bibinfo {author} {\bibfnamefont {C.}~\bibnamefont {Chamon}},\ }\href
  {\doibase 10.1103/PhysRevB.82.144513} {\bibfield  {journal} {\bibinfo
  {journal} {Phys. Rev. B}\ }\textbf {\bibinfo {volume} {82}},\ \bibinfo
  {pages} {144513} (\bibinfo {year} {2010})}\BibitemShut {NoStop}%
\bibitem [{\citenamefont {Quelle}\ \emph {et~al.}(2016)\citenamefont {Quelle},
  \citenamefont {Smith}, \citenamefont {Kvorning},\ and\ \citenamefont
  {Hansson}}]{kvorning}%
  \BibitemOpen
  \bibfield  {author} {\bibinfo {author} {\bibfnamefont {A.}~\bibnamefont
  {Quelle}}, \bibinfo {author} {\bibfnamefont {C.~M.}\ \bibnamefont {Smith}},
  \bibinfo {author} {\bibfnamefont {T.}~\bibnamefont {Kvorning}}, \ and\
  \bibinfo {author} {\bibfnamefont {T.~H.}\ \bibnamefont {Hansson}},\ }\href
  {\doibase 10.1103/PhysRevB.94.125137} {\bibfield  {journal} {\bibinfo
  {journal} {Phys. Rev. B}\ }\textbf {\bibinfo {volume} {94}},\ \bibinfo
  {pages} {125137} (\bibinfo {year} {2016})}\BibitemShut {NoStop}%
\end{thebibliography}%

\end{document}